\documentclass[]{aa}

\usepackage{graphicx, xspace, lscape}
\usepackage{amsmath}
\usepackage{longtable}
\usepackage{float}

\bibpunct{(}{)}{;}{a}{}{,}
\def\lsim{\mathrel{\rlap{\lower4pt\hbox{\hskip1pt$\sim$}}
    \raise1pt\hbox{$<$}}}
\def\gsim{\mathrel{\rlap{\lower4pt\hbox{\hskip1pt$\sim$}}\raise1pt\hbox{$>$}}}
\def\ang{\hbox{$\text{\AA}$}\xspace}
\def\kms{$\mbox{km~s}^{-1}$\xspace}
\def\lam{$\lambda$}
\def\lamlam{$\lambda\lambda$}
\def\Msun{$M_\odot$\xspace}
\def\y{$\bullet$}
\begin{document}
   \title{A VLT/FLAMES survey for massive binaries in Westerlund~1: VIII. Binary Systems and Orbital Parameters}

   \author{B.~W.~Ritchie\inst{1,2} \and J.~S.~Clark\inst{1, \thanks{In Memoriam: This work is dedicated to the memory of Dr. Simon Clark, who tragically passed away while the paper was in preparation. His friendship and passion for Astronomy are deeply missed by all who worked with him.}} \and I.~Negueruela\inst{3} \and F.~Najarro\inst{4}}

   \institute{
      Department of Physics and Astronomy, The Open University, Walton Hall, Milton Keynes, MK7 6AA, UK.\\
      \email{ben.ritchie@open.ac.uk}
   \and
      Lockheed Martin Integrated Systems, Building 7000, Langstone, Hampshire, PO9 1SA, UK.
   \and
      Departamento de F\'{\i}sica Aplicada, Facultad de Ciencias, Universidad de Alicante, Carretera de San Vicente s/n, E03690, San Vicente del Raspeig, Spain
   \and
      Centro de Astrobiolog\'{i}a, CSIC-INTA, Ctra de Torrej\'{o}n a Ajalvir km 4, E-28850 Torrej\'{o}n de Ardoz, Madrid, Spain.
   }

   \date{Received 18 October 2021 / Accepted 24 November 2021}

   \abstract
   {The galactic cluster Westerlund~1 contains a rich population of evolved, massive stars. A high
   binary fraction has been inferred from previous multiwavelength observations.}
   {We use multi-epoch spectroscopy of a large sample of early-type stars in Westerlund~1 to
   identify new binaries and binary candidates in the cluster.}
   {VLT/FLAMES was used with the GIRAFFE spectrograph in HR21 mode to obtain spectra of $\sim$100 OB
   stars over a 14-month baseline in 2008 and 2009, supplemented with follow-up observations in 2011
   and 2013. Radial velocities were obtained from strong Paschen series absorption lines in the
   $I$-band.}
   {We identify 20 new OB I--III binaries, a WN9h: binary, and a WC9d binary, greatly increasing the
   number of directly confirmed binary systems in Westerlund~1, while 12 O9--9.5 Iab--III stars are
   identified as candidate binaries based on radial velocity changes that are inconsistent with
   photospheric variability. The 173.9~day SB1 \object{W1030} represents the first longer-period
   system identified in the cluster, while the determination of a 53.95~day period for \object{W44}
   (\object{WR~L}) makes it the first Wolf-Rayet binary in Westerlund~1 with a confirmed orbital
   period greater than ten days. Our results suggest the binary fraction in the OB population is at
   least $\sim$40$\%$, and may be significantly higher.}
   {These results demonstrate that binary systems can be effectively identified in the population of
   OB I--III stars evolving off the main sequence in Westerlund~1. Future multi-epoch surveys will
   be able to fully characterise this population.}

  \titlerunning{OB binaries in Wd1}
  \maketitle

\section{Introduction}

In recent years a number of large multi-epoch spectroscopic radial velocity (RV) surveys have been
carried out to characterise the high-mass binary populations in young, massive clusters
($M\gsim10^4$\Msun, ages of a few Myr), focusing on both Galactic clusters and the Small and Large
Magellanic clouds (e.g. \citealt{evans08, bosch09, ritchie09a, clark15, almeida, lohr18, evans20},
and references therein). These studies are motivated by the high binary fraction amongst massive
stars \citep{sana12} and the critical role binary interaction plays throughout the late-stage
evolution and ultimate core-collapse of such systems \citep{langer03, schneider15}, with their
relativistic remnants powering a range of high-energy phenomena such as $\gamma$-ray bursts,
high-mass X-ray and $\gamma$-ray binaries, and, eventually, the generation of gravitational waves
from the inspiral and coalescence of neutron stars and black holes. 

In this paper, we report the full results of a study of OB stars in the very massive galactic
cluster Westerlund~1 (hereafter Wd1; \citealt{w61, cncg05, neg10, clark20}) with the Fibre Large
Array Multi Element Spectrograph (FLAMES; \citealt{pasq02}), based on observations obtained in
2008--2009 and supplemented with additional FLAMES observations obtained during 2011 and 2013.
Although heavily reddened, the luminous stellar population of Wd1 can be observed in the $R$- and
$I$-bands, revealing an apparently-coeval population of OB stars\footnote{Wd1 contains a large
population of $\sim$O9~II-III stars, implying a MS turn-off around $\sim$O8~V \citep{clark20}, but
to date no examples of the early-B~II-III stars that would trace an older stellar population
evolving off the MS have been observed; in the $I$-band, such objects would be easily
distinguished from the late-O population by the presence of He~I and N~I absorption lines and the
absence of C~III \citep{neg10}.} evolving off the main sequence (MS), along with a rich population
of post-MS objects that include blue supergiants (BSGs), yellow hypergiants (YHGs), red supergiants
(RSGs), and Wolf-Rayet (WR) stars \citep{clark10, clark20}. A photometric study of the stellar
population by \cite{bonanos07} found four eclipsing binary systems with periods of less than ten
days and a similar number of systems with ellipsoidal variations in their light curves suggestive of
binarity, while observations at radio \citep{dougherty}, infra-red \citep{crowther06} and X-ray
wavelengths \citep{skinner,clark08,clark19b} also suggest that Wd1 is binary-rich, although these
wavelengths probe secondary indicators of binarity from wind interaction: X-ray emission from
colliding winds \citep{stevens92}, an infra-red excess due to hot dust formed in a wind-interaction
zone \citep{williams}, or non-thermal radio spectra \citep{debecker}. 

Short-period binaries in Wd1 are expected to undergo Roche-lobe overflow as the primary evolves off
the MS \citep{langer03,petrovic} and will become WR+O binaries without ever reaching the B--F
hypergiant phase (cf. \object{W13}, \object{W72}/{WR~A}, \object{WR~B}, and \object{W239}/{WR~F};
\citealt{bonanos07,ritchie10,clark11,kb12}). Wd1 is therefore expected to have a cool-phase
population of single stars or very widely-separated, non-interacting binaries, while binary-mediated
evolution in short-period systems is reflected in the very high binary fraction in the Wolf-Rayet
population ($\gsim70\%$; \citealt{crowther06}; \citealt{clark08}). This theoretical model is
supported by the apparent lack of binaries amongst the population of very luminous mid-B to F
hypergiants: to date, the only candidate binary in the cool-phase population is the LBV
\object{W243}, which has a potential interferometric companion \citep{mahy21}, an X-ray excess
consistent with the presence of a late-O supergiant \citep{clark19b}, and a complex H and He~I
emission-line spectrum that suggests a hot secondary is ionising the wind of the cool-phase primary
\citep{ritchie09b}. The presence of high-amplitude pulsational instabilities \citep{ritchie09a,
clark10} and the large orbital separation required to accommodate a B--F hypergiant make RV studies
of this population extraordinarily challenging, but if any of these cool-phase hypergiants are
binary, then they must have such a large separation that the primary can evolve across the
Hertzsprung-Russell diagram as an effectively-isolated object, and the most luminous objects in Wd1
are unlikely to be representative of the unevolved binary population in the cluster. In this paper
we therefore examine a dataset of more than a hundred O9--B2 stars with masses $\sim$25--35\Msun
\citep{clark20} that are just starting to evolve away from the MS. Unless their initial periods are
extremely short, binaries in this population are not expected to have begun to interact
\citep{wellstein01}, and these systems provide direct insight into the nature of the binary
population of Wd1 before the evolutionary pathways for single and multiple stars start to diverge.

Initial results from five epochs of data obtained in 2008 were presented in \cite{ritchie09a},
hereafter Paper~I, which identified two short-period binaries (\object{W43a} and \object{W1065}) and
two candidates (\object{W2a} and \object{W232}). Further results drawn from a 14-month dataset
obtained in 2008--2009 include an orbital solution for the 23+35\Msun double-lined eclipsing binary
\object{W13} \citep{ritchie10}, and the identification of a $5.05$~day orbital period in the X-ray
luminous WC9d+O binary \object{W239}/{WR~F} \citep{clark11} which likely includes a third massive
component responsible for episodic formation of hot dust. The ultimate endpoint of the binary
evolution channel are objects like the B1.5~Ia$^{+}$/WNVL hypergiant \object{W5}, identified as a
potential remnant of a binary system disrupted by the primary supernova event \citep{clark14}, while
other papers in this series have focused on complex evolved systems such as the blue stragglers
\object{W27} and \object{W30a} (O7--8~Ia$^+$ and O4--5~Ia$^+$ respectively; \citealt{clark19a}) and
the WN9h: binary \object{W44}/\object{WR~L} (Ritchie~et~al., in~prep.). Properties of X-ray bright
massive cluster members are discussed by \cite{clark19b}, hereafter Paper~II, and a cluster census,
combining the observations discussed here with other archival observations, is presented in
\cite{clark20}, hereafter Paper~III.  

\section{Observations \& data reduction}
\label{sec:obs_data}
\begin{table}[!htpb]
  \caption{Census of 151 targets included in the FLAMES datasets referenced in this paper, broken down by spectral type from Paper~III. The O9--9.5~III category  includes objects classified as candidate O~binaries based on $I$-band spectral morphology.}
  \label{tab:spectype}
  \begin{center}
  \begin{tabular}{ll}
  Spectral Type & Count \\
  \hline\hline
  &\\
  O4-7~Ia$^+$          & 2$^a$\\ 
  O9--9.5~III~/~O~bin? & 50 \\ 
  O9--9.5~I-II         & 27 \\ 
  B0--B1.5~Ia--Ib      & 23 \\ 
  B2--B4~Ia            & 5\\ 
  B2--B9~Ia$^+$        & 4$^{b,c}$\\ 
  A--F~Ia$^+$          & 5$^{b}$\\ 
  M0--4~Ia             & 2$^b$ \\ 
  \hline
  &\\
  LBV                     & 1$^d$ \\
  sgB[e]                  & 1$^e$ \\
  Oe~/~Be                 & 1$^c$ \\
  B0.5--1.5~Ia$^+$ / WNVL & 2$^{f,g}$\\ 
  WN8--9                  & 2 \\ 
  WN5--7                  & 9 \\
  WC8--9                  & 8 \\
  \hline
  &\\
  Field stars     & 9$^*$\\
\end{tabular}
\end{center}
$^a$\cite{clark19a}, $^b$\cite{clark10}, $^c$Paper~III, $^d$\cite{ritchie09b}, $^e$\cite{clark13}, $^f$\cite{ritchie10}, $^g$\cite{clark14}.
$^*$One very faint `field' object may be a cluster member, but membership cannot be confirmed. 
\end{table}

The VLT/FLAMES survey for massive binaries in Westerlund~1 carried out observations in 2008, 2009,
2011, and 2013, using the FLAMES-GIRAFFE multi-fibre spectrograph on VLT UT2 \textit{Kueyen} with
setup HR21 covering the 8484-9001\AA~range with $R=\lambda/\Delta\lambda \sim 16200$. The
FLAMES-GIRAFFE pipeline was used to bias subtract, flat-field, and wavelength calibrate the data,
while subsequent processing made use of IRAF for the extraction of individual spectra as well as
rectification and heliocentric velocity correction. Full details of target selection, fibre
prioritisation, and data reduction are given in Paper~I, with additional discussion in Paper~III. A
summary of the full set of targets by spectral type is given in Table~\ref{tab:spectype}. 

To briefly summarise target selection, the 2008 and 2009 observations focused on four target fields.
A \textit{bright} field, containing 17 spectroscopically-selected targets (14 OB supergiants, two
A--F hypergiants, and a WC9d Wolf-Rayet) and five bright photometrically-selected
targets\footnote{Identified in Paper~I as four O9.5--B0~Iab/b cluster members and one field
M2~II/III star.}, was observed on eleven epochs, while three \textit{faint} fields containing 17
spectroscopically-confirmed cluster members (16 OB supergiants and a WC8 Wolf-Rayet) and 63
photometrically-selected candidates\footnote{Identified in Paper~III as 55 O9~III--B0~Ia cluster
members, one Oe/Be cluster member, and seven field stars.} were observed on seven epochs
(\textit{faint1}) or six epochs (\textit{faint2}, \textit{faint3}). Observations are summarised in
Table~\ref{tab:observations}. A few targets are present in more than one list; duplicates in the
\textit{faint} lists were not accounted for in the description of targets in Paper~I but have been
removed here, leading to slight differences in the number of targets reported. The 2008--2009
dataset was supplemented with additional FLAMES observations obtained in 2011 and 2013, which are
summarised in Table~\ref{tab:additional_observations}. Observations in 2011 were primarily focused
on two sets of Wolf-Rayet targets (identified as \textit{WN} and \textit{WC} fields), but spare
fibres were also allocated to targets of interest in the OB stellar population. 2013 observations
focused on three sets of targets, and provided low (\textit{L1\dots3}) and high (\textit{H1\dots3})
resolution follow-up observations of objects of interest from the 2008--2009 dataset, along with
initial observations of a number of OB targets that had not been observed in previous FLAMES
campaigns. The majority of the 2011 and 2013 observations were obtained using setup HR21 and were
reduced in the same manner as earlier observations, and only the HR21 setup was used to measure RVs.
Observations obtained in other FLAMES modes were used to examine the $R$-band spectra of certain
targets; these are described in the text when used. 

In addition to the core 2008--2013 dataset, archival observations were used to further investigate a
number of targets. Low-resolution FLAMES observations from 2005 focused on a broad set of OB
supergiant and late-type targets (labeled as \textit{SG} fields in
Table~\ref{tab:additional_observations}), and are described in more detail by \cite{clark10}. In
addition, $R$-band spectra of a number of our targets were acquired in June 2004 using VLT/FORS2 in
longslit and mask exchange unit (MXU) modes \citep{neg10}. 

\begin{table}[!htpb]
\caption{Journal of observations during 2008--2009. All observations are in FLAMES~HR21 mode.}
\label{tab:observations}
\begin{center}
\begin{tabular}{llcll}
Date       & MJD$^{a}$   & $\Delta$t$^{b}$ & Targets & Int. \\
\hline\hline
&&&&\\
20/06/2008 & 54637.1895 & --              & Faint 1 & 3$\times$895s\\
           & 54637.2304 & --              & Faint 2 & 3$\times$895s\\
&&&&\\
29/06/2008 & 54646.1846 & --              & Bright  & 2$\times$600s\\
           & 54646.2239 & --              & Faint 3 & 3$\times$895s\\
&&&&\\
18/07/2008 & 54665.0356 & 18.85           & Bright  & 2$\times$600s\\
           & 54665.0651 & 18.84           & Faint 3 & 3$\times$895s\\
           & 54665.1041 & 27.87           & Faint 2 & 3$\times$895s\\
           & 54665.1439 & 27.95           & Faint 1 & 3$\times$895s\\
&&&&\\
24/07/2008 & 54671.1343 &  6.10           & Bright  & 2$\times$600s\\
&&&&\\
14/08/2008 & 54692.0423 & 20.91           & Bright  & 2$\times$600s\\
           & 54692.0731 & 27.01           & Faint 3 & 3$\times$895s\\
&&&&\\
17/08/2008 & 54695.0941 & 29.99           & Faint 2 & 3$\times$895s\\
           & 54695.1328 & 29.99           & Faint 1 & 3$\times$895s\\
&&&&\\
04/09/2008 & 54713.0107 & 20.97           & Bright  & 2$\times$500s\\
&&&&\\
15/09/2008 & 54724.0818 & 11.07           & Bright  & 2$\times$500s\\
&&&&\\
17/09/2008 & 54726.0148 & 33.94           & Faint 3 & 3$\times$895s\\
           & 54726.0532 & 30.96           & Faint 2 & 3$\times$895s\\
&&&&\\
19/09/2008 & 54728.0245 & 32.89           & Faint 1 & 3$\times$895s\\
           & 54728.0554 & 3.97            & Bright  & 600,700s\\
&&&&\\
25/09/2008 & 54734.0613 & 6.01            & Bright  & 2$\times$600s\\
&&&&\\
14/05/2009 & 54965.1768 & 231.12          & Bright  & 2$\times$600s\\
           & 54965.2061 & 239.19          & Faint 3 & 3$\times$865s\\
           & 54965.2448 & 239.19          & Faint 2 & 3$\times$865s\\
           & 54965.2827 & 237.26          & Faint 1 & 3$\times$865s\\
&&&&\\
18/05/2009 & 54969.3198 & 4.14            & Bright  & 2$\times$600s\\
           & 54969.3478 & 4.07            & Faint 1 & 3$\times$865s\\
&&&&\\
18/08/2009 & 55061.0310 & 95.82           & Faint 3 & 3$\times$865s\\
           & 55061.0680 & 95.82           & Faint 2 & 3$\times$865s\\
&&&&\\
20/08/2009 & 55063.0290 & 93.68           & Faint 1 & 3$\times$865s\\
           & 55063.0575 & 93.74           & Bright  & 2$\times$600s\\
&&&&\\
\hline
\end{tabular}
\end{center}
$^{a}$The modified Julian day (MJD) is given at the midpoint of integrations.\\
$^{b}$The time (in days) since the target field was last observed.\\
\end{table}

\begin{table}[!htpb]
\caption{Journal of supplemental FLAMES observations during 2005, 2011, and 2013. HR21 mode was used unless noted.}
\label{tab:additional_observations}
\begin{center}
\begin{tabular}{llcll}
Date       & MJD$^{a}$   & $\Delta$t$^{b}$ & Targets & Int. \\
\hline\hline
&&&&\\
23/03/2005 & 53454.31865 & --             & SG$^{c}$ & 2$\times$877s\\
           & 53454.34250 & --             & SG$^{d}$ & 2$\times$600s\\
           & 53454.35167 & --             & SG$^{e}$ & 2$\times$100s\\
&&&&\\
29/05/2005 & 53519.25658 & 64.91          & SG$^{d}$ & 2$\times$530s\\
           & 53519.26506 & 64.91          & SG$^{e}$ & 2$\times$100s\\
&&&&\\
13/07/2005 & 53564.10112 & 109.78         & SG$^{c}$ & 2$\times$877s\\
           & 53564.12375 & 44.87          & SG$^{d}$ & 2$\times$600s\\
           & 53564.13294 & 44.87          & SG$^{e}$ & 2$\times$100s\\
\hline 
&&&&\\
17/04/2011 & 55668.2404 & --              & WC & 3$\times$865s\\
&&&&\\
26/04/2011 & 55677.2063 & --              & WN$^f$ & 3$\times$865s\\
&&&&\\
20/05/2011 & 55701.1855 & 23.98           & WN & 3$\times$865s\\
&&&&\\
22/05/2011 & 55703.1494 & 34.91           & WC & 3$\times$865s\\
&&&&\\
24/06/2011 & 55736.9864 & 35.80           & WN & 3$\times$865s\\
&&&&\\
25/06/2011 & 55737.0293 & 33.88           & WC & 3$\times$865s\\
&&&&\\
30/08/2011 & 55804.0037 & --              & SG & 3$\times$895s\\
\hline
&&&&\\
12/04/2013 & 56394.2703 & --              & L1$^d$ & 3$\times$894s\\
           & 56394.3156 & --              & L2$^d$ & 3$\times$894s\\
           & 56394.3598 & --              & L3$^d$ & 3$\times$894s\\
&&&&\\
20/05/2013 & 56432.1406 & 37.87           & L1$^d$ & 3$\times$894s\\
&&&&\\
10/07/2013 & 56483.1993 & --              & H1 & 3$\times$815s\\
           & 56483.2356 & --              & H2 & 3$\times$815s\\
           & 56483.2713 & --              & H3 & 3$\times$815s\\
&&&&\\
21/07/2013 & 56494.0829 & 10.88           & H1 & 3$\times$815s\\
           & 56494.1207 & 10.89           & H2 & 3$\times$815s\\
           & 56494.1999 & 10.93           & H3 & 3$\times$815s\\
&&&&\\
23/07/2013 & 56496.0887 & 2.01            & H1 & 9$\times$815s\\
&&&&\\
24/07/2013 & 56497.1969 & 3.08            & H2 & 3$\times$815s\\
&&&&\\
04/08/2013 & 56508.1094 & 13.91           & H3 & 3$\times$815s\\
           & 56508.1730 & 12.08           & H1 & 3$\times$815s\\
&&&&\\
07/08/2013 & 56511.1590 & 13.97           & H2 & 2$\times$815s\\
&&&&\\
08/08/2013 & 56512.2129 & 1.05            & H2 & 3$\times$815s\\
&&&&\\
\hline
\end{tabular}
\end{center}
$^{a}$The modified Julian day (MJD) is given at the midpoint of integrations.\\
$^{b}$The time (in days) since the target field was last observed.\\
$^{c}$LR4 mode.\\ 
$^{d}$LR6 mode.\\ 
$^{e}$LR8 mode.\\ 
$^{f}$HR16 mode.
\end{table}

\section{Analysis}
\subsection{Radial velocity measurement}
\label{sec:rv}

\begin{figure}
\begin{center}
\resizebox{\hsize}{!}{\includegraphics{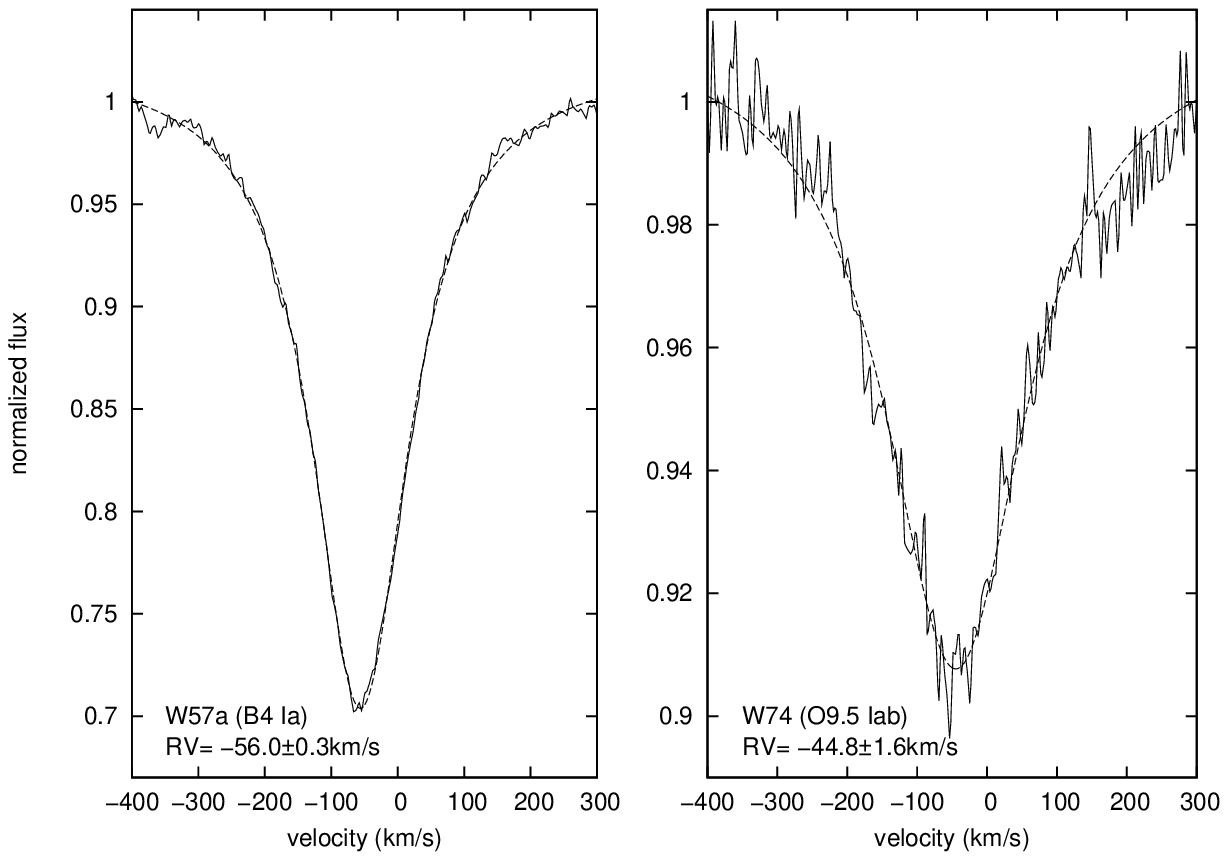}}
\resizebox{\hsize}{!}{\includegraphics{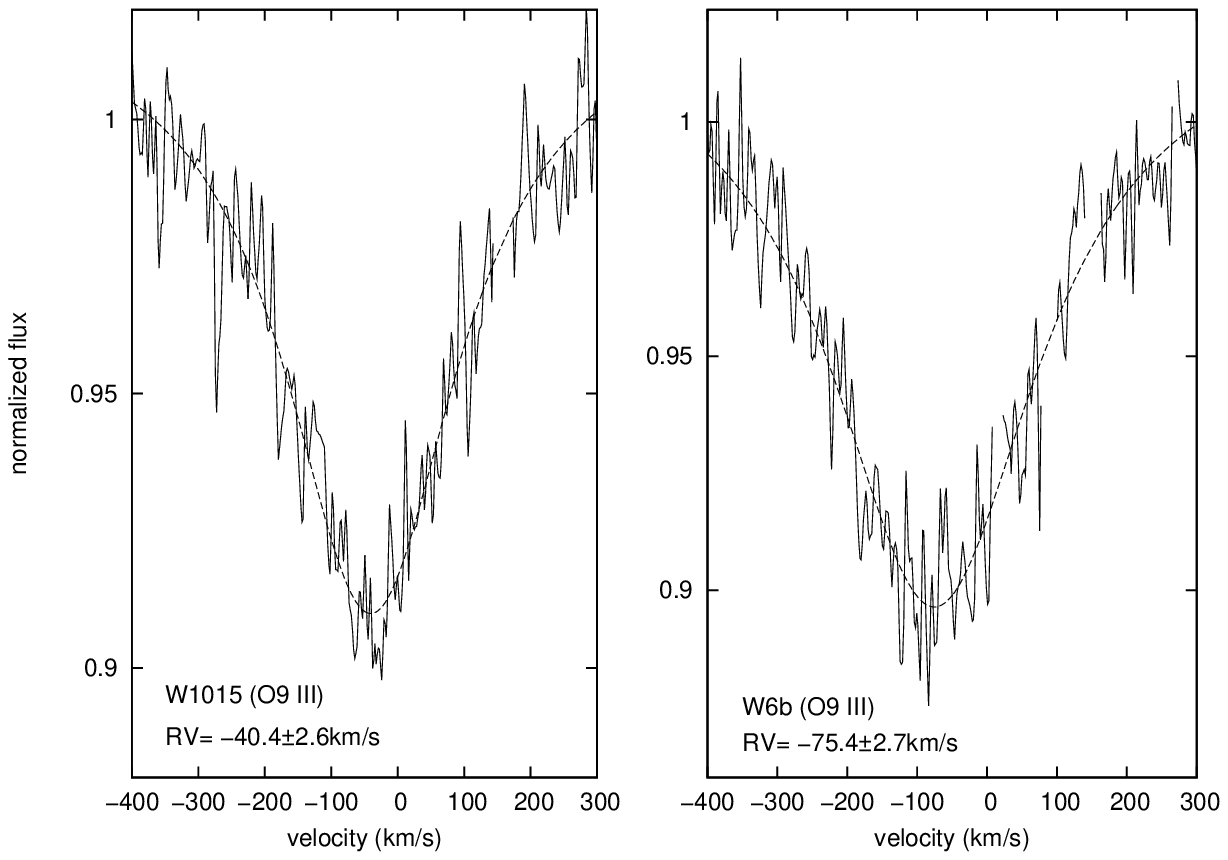}}
\caption{Example fits to the Pa-11 line at a variety of $S$/$N$ representative of the targets included in this programme. Note that \object{W6b} is a single-lined binary.}
\label{fig:measure}
\end{center}
\end{figure}
RVs were measured by fitting Lorentzian profiles to the strong Paschen series lines in the
8500--9000\ang range covered by the FLAMES HR21 setup, using the Levenberg-Marquardt nonlinear
least-squares method; example fits to lines with both high and low signal-to-noise ratios ($S$/$N$)
are shown in Figure~\ref{fig:measure}, and summary results for all objects are given in
Table~\ref{tab:fullresults}. Fits were made from $-500$ to $+500$\kms relative to the line rest
wavelength, which ensures that at low $S$/$N$ the derived RV is not significantly affected by noise
or sky residuals overlapping the line core. Exceptions were made in the following cases:
\begin{itemize}
  \item When fitting the Pa-11~\lam8862 line in B supergiants the fit was only extended to $-400$\kms to avoid the influence of the He~I~\lam8845 line bluewards of Pa-11. In very high $S$/$N$ spectra a weak DIB from the C$_2$ Phillips (2-0) system is visible on the blue wing of Pa-11 (see \object{W57a} in Figure~\ref{fig:measure}), but does not affect the fit.
  \item A broad DIB overlaps Pa-13~\lam8665, strongly attenuating the blue wing of the line, and the fit was limited to $\pm$200\kms from the line centre.
  \item In very broad-lined O-type stars the line width can exceed $\pm$500\kms, and the fitting region was extended to fully accommodate the line profile.
\end{itemize}
In a number of cases, targets were included in more than one target list and were observed twice
(e.g. \object{W55}, \object{W71}, or \object{W1036}), or even three times (\object{W232},
\object{W1065}) on the same night. In all but one case measured RVs are in agreement to within the
respective fitting errors, and typically agree to within $\sim$1\kms, suggesting that our RV
measurements are not significantly affected by transient noise or sky residuals; \object{W232} is
the sole exception, and is discussed further in Section~\ref{sec:w232}. A strong, well-defined DIB
at $8620\ang$ \citep{damineli16} serves as a check for zero-point errors in the wavelength
calibration of individual spectra. 

While the Pa-12~\lam8750 line was included in the analysis in Paper~I, subsequent examination of the
effect of the C$_2$ Phillips (2--0) band system on the redwards flank of the line has led to this
line being excluded from this analysis, as these interstellar features lead to a broadening and
systematic shift redwards by $\sim$10\kms relative to Pa-11 and Pa-13 (see also discussion in
\citealt{ritchie10}). While the increased fitting error caused by the asymmetric Pa-12 line profile
reduces the impact of this offset in the RVs reported in Paper~I, we find an improved RV measurement
if the line is omitted completely, and in early B objects Pa-11 and Pa-13\dots15 in are excellent
agreement. In high $S$/$N$ spectra a bluewards shift in the Pa-16~\lam8502 line from blending with
C~III~\lam8500 can be seen in objects as late as B0.5--1~Ia, and this line is also consequently
excluded. However, most of the targets considered in this paper come from a population of
lower-luminosity stars with spectral types earlier than B0. The higher Paschen series lines are
fading rapidly in these objects, and RVs were measured using the Pa-11 and Pa-13 lines alone. The
decreasing $S$/$N$, weakening line strength, and increasing line width towards lower luminosity
classes leads to a systematic increase in RV measurement errors towards earlier spectral types:
while errors internal to the Levenberg-Marquardt fit are generally below 1\kms for all Paschen lines
in the most luminous objects in our target list, which have $S$/$N>$100, at O9.5~II mean fitting
errors are generally around 3\kms for Pa-11 and $\gg$5\kms for Pa-13 and Pa-14, leading to an
error-weighted RV dominated by the measurement of Pa-11. We also note that a subset of faint objects
(with $I\ge14.5$) have very broad, weak Paschen series lines and significantly higher fitting
errors; these were proposed as heavily-blended SB2s in Paper~III, and we briefly return to these
objects in Section~\ref{sec:broad}.

\subsection{Sky emission lines}
\label{sec:sky}
\begin{figure}
  \begin{center}
  \resizebox{\hsize}{!}{\includegraphics{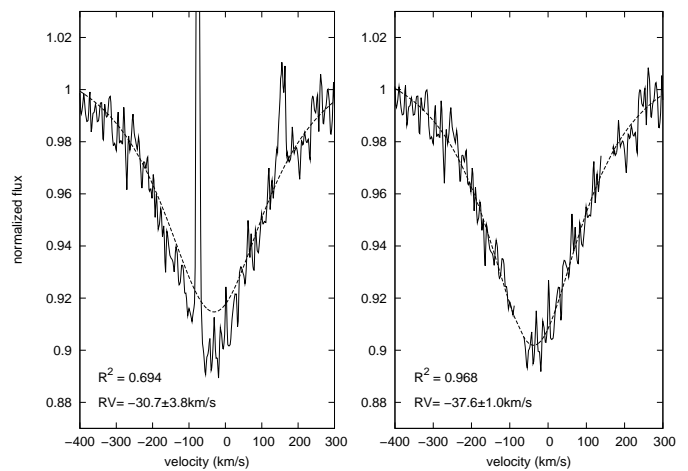}}
  \caption{Effect of unsubtracted sky lines on RV determination.
  (\textit{Left Panel}) Pa-11 line profile for \object{W17} (20/06/2008)
  with unsubtracted lines at $-75$\kms and $+160$\kms. (\textit{Right Panel}) the same Pa-11 line after the sky lines were removed. In each panel, the RV and goodness-of-fit metric $R^2$ derived from the Levenberg-Marquardt method are listed, demonstrating the reduced error and improved fit following sky line removal.}
  \label{fig:skylines}
  \end{center}
\end{figure}

Sky emission lines that are not fully removed by sky spectra subtraction are often problematic,
especially for the lower-luminosity targets in our dataset. In the majority of cases these lines do
not fall near the Paschen series absorption lines used for RV measurement, but when these features
overlap the resultant fit may be a poor match to the line core: an example is shown in the left
panel of Figure~\ref{fig:skylines}, with a strong sky residual reaching $\sim$20\% above the
continuum level falling $\sim$40\kms bluewards of the line centre. Such features are easily
identified as a significant departure from the local moving average, and we take the deliberately
simplistic approach of locating the peak of the sky emission line and excluding the five data bins
on either side, with testing showing that more complex removal does not result in appreciable
improvement. The right panel of Figure~\ref{fig:skylines} shows the improved fit after sky line
removal. 

\subsection{Binary candidate identification}
\label{sec:cands}
\begin{figure}
  \begin{center}
    \resizebox{\hsize}{!}{\includegraphics{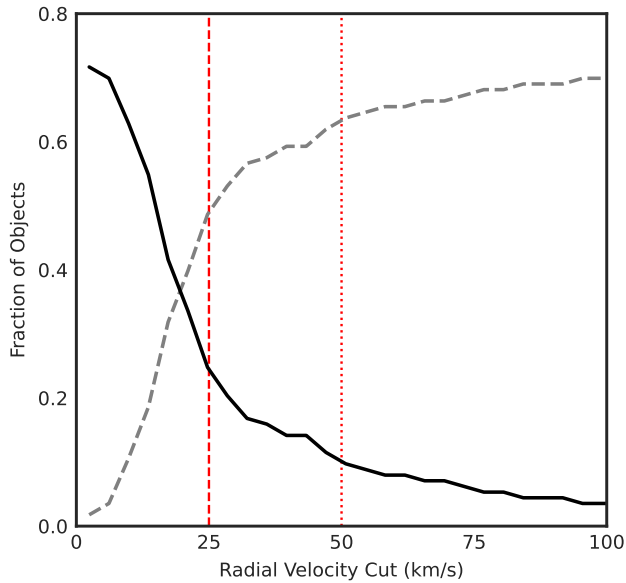}}
    \caption{Fraction of systems passing a RV cut at $C$~\kms to separate binaries from pulsational variables. The solid line marks the fraction identified as binaries ($\Delta$RV$>C$), while the dashed line identifies the fraction identified as variables ($\Delta$RV$<C$). The vertical lines mark the 25\kms (dashes) and 50\kms (dots) thresholds discussed in the text.}
    \label{fig:rvs}
  \end{center}
\end{figure}

As discussed in Paper~I, RV variability is not necessarily orbital, and may also result from wind
effects and radial and/or non-radial pulsations (see also \citealt{sana13} and
Lohr~et~al.~in~prep.). Taking the LBV \object{W243} (A3~Ia$^+$) as an example of a high-amplitude
pulsator, 20 epochs of data display RVs spanning $\sim$24\kms with variability of $\sim$1--2\kms on
a timescale of days (individual fitting errors $<1$\kms). No periodicity is apparent, and the sheer
size of W243 ($R_* \sim 420R_\odot$; \citealt{ritchie09b}) precludes a binary interpretation for the
rapid RV variability we observe. Luminous B~supergiants and A--F hypergiants display variability of
similar magnitude and cadence \citep{clark10}. We therefore follow \cite{sana13} and examine the
maximum significance of RV differences between any two epochs of observation
\begin{equation}
  \label{eq:cut}
  \sigma_{\textrm{det}}=\textrm{max}\left(\frac{|v_i-v_j|}{\sqrt{\sigma_i^2 + \sigma_j^2}}\right)
\end{equation}
and require that at least one pair of observations display $\Delta\textrm{RV}=|v_i-v_j|$ that
exceeds a minimum amplitude threshold $C$. As apparently non-orbital variability in measured RVs can
exceed $\sim$20\kms in the luminous evolved population of early--mid~B supergiants in Wd1, we adopt
$C>25$\kms and $\sigma_{\textrm{det}}>4$ as indicative of a likely binary nature: failure to meet
this clip level does not preclude binarity, but means that we cannot separate orbital motion from
other effects with the available data. The classification of targets as spectroscopic binaries is
clearly dependent on the choice of velocity cut, and Figure~\ref{fig:rvs} shows the fraction of
objects selected as candidate binaries or pulsational variables as the cut-off velocity is
increased. A break in the distribution is seen at $\sim$28\kms, suggesting that this may mark the
maximum amplitude of wind and/or pulsational effects. This is in good agreement with the 25--30\kms
threshold for OB supergiants suggested by \cite{sd20}. Although this is slightly above our adopted
25\kms threshold, potential photospheric variability in this range is limited to a few
highly-luminous B supergiants and increasing the threshold further would start to eliminate a number
of systems that are clearly \textit{bona fide} binaries. A simple extrapolation of the slope above
$\sim$40\kms towards lower RVs would suggest that roughly $\sim$10\% of our total targets may be
low-amplitude binaries that cannot be distinguished from a larger population of pulsational
variables. We return to this topic in Section~\ref{sec:discussion}. 

\subsection{Period search}
\label{sec:period}
\begin{figure}
  \begin{center}
    \resizebox{\hsize}{!}{\includegraphics{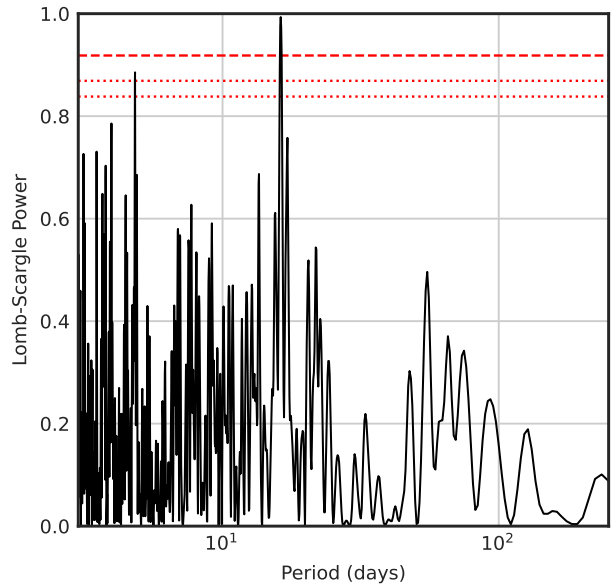}}
    \caption{Lomb-Scargle Periodogram for \object{W43a}. Horizontal lines mark (from top to bottom) \cite{baluev} $1\%$, $5\%$, and $10\%$ false alarm probabilities. The strongest peak is at 16.241~days, and the minimum string length (not shown) occurs at 16.243~days.} 
    \label{fig:lombscargle}
  \end{center}
\end{figure}

To search for periodicities in the RV data of binary candidates, we first compute the Lomb-Scargle
periodogram \citep{lomb, scargle}: an example is shown in Figure~\ref{fig:lombscargle}. We take
250~days as the upper limit for the search, and use the target's spectral type as a guide to the
lower limit, adopting 1~day for O9~III, 2~days for O9--9.5~I--II, 3~days for B0--B1.5~Ia, and ten
days for objects of spectral type B2--4~Ia; in each case, we consider shorter periods to be
unphysical, as they require an orbital separation  incompatible with the spatial extent of a
(non-interacting) primary (cf. \citealt{martins}). We then carry out a string-length minimisation
search \citep{dworetsky} to confirm any identified period. If the two methods are in agreement, the
final determination of the orbital period is carried out using the Li\`{e}ge Orbital Solution
Package\footnote{https://stsci.edu/$\sim$hsana/losp.html} (hereafter LOSP), with non-zero
eccentricity permitted. We regard an orbital period as secure if:
\begin{enumerate}
  \item The strongest peak in the Lomb-Scargle periodogram has a \cite{baluev} false alarm probability $<1\%$.
  \item The periodogram, string-length search, and LOSP period search identify the same orbital period.
  \item No alternative periods with similar peak strengths exist. 
\end{enumerate}
If the false alarm probability is in the range $1-5\%$ then we refer to the solution as a
\textit{candidate} period provided that the other conditions are met. In some cases we discuss
possible orbital periods that do not meet these criteria, noting that these \textit{tentative}
identifications must be confirmed with future observations. 

\subsection{Orbital parameters}
\label{sec:orbital}
The systemic velocity $\gamma$, semi-amplitude $K$, eccentricity $e$, and longitude of periastron
$\omega$ were determined using LOSP, fixing the orbital period $P$ at the value determined by the
period search discussed in Section~\ref{sec:period}. As an independent check, we then repeated the
analysis using THE~JOKER \citep{pw17}, a custom Monte~Carlo sampler for sparse RV datasets: in all
cases, the two methods produced orbital solutions that were in agreement to within their respective
errors. Orbital solutions from LOSP are discussed in Section~\ref{sec:individual} and summarised in
Table~\ref{tab:results}. 

We note that many objects only have limited sampling, and in the majority of cases orbital solutions
should be regarded as provisional: for this reason, we do not attempt more detailed analysis of
aggregate parameters at this time (e.g. the distribution of orbital eccentricities). In particular,
poor sampling of the extremes of the RV curve or (in eccentric systems) the periastron passage may
lead to significant inaccuracies in orbital parameters. Comparison of our results with an initial
analysis based on the 2008--2009 dataset alone suggests that additional data do not
\textit{substantially} affect our conclusions; further observations have not ruled out any targets
that were identified as short-period binaries in the initial analysis. However, additional data
often led to small changes in the determinations of semi-amplitude, period, and eccentricity, and we
expect that future observations will continue to refine the results presented here.  

\section{Binary Identification}
\begin{figure}
  \begin{center}
    \resizebox{\hsize}{!}{\includegraphics{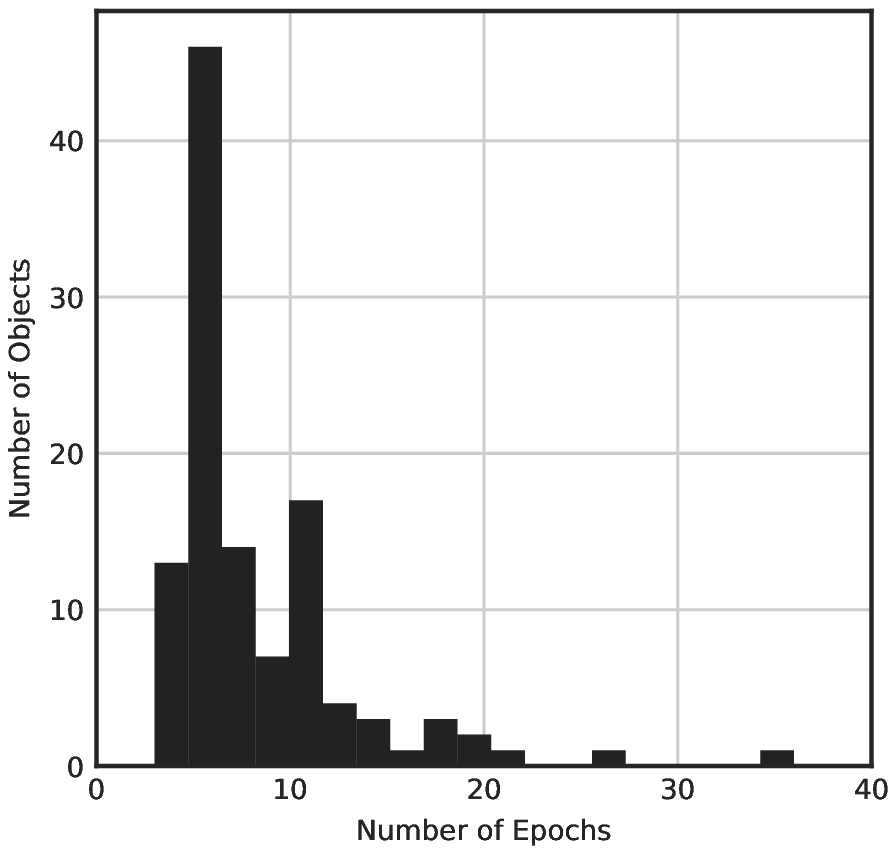}}
    \resizebox{\hsize}{!}{\includegraphics{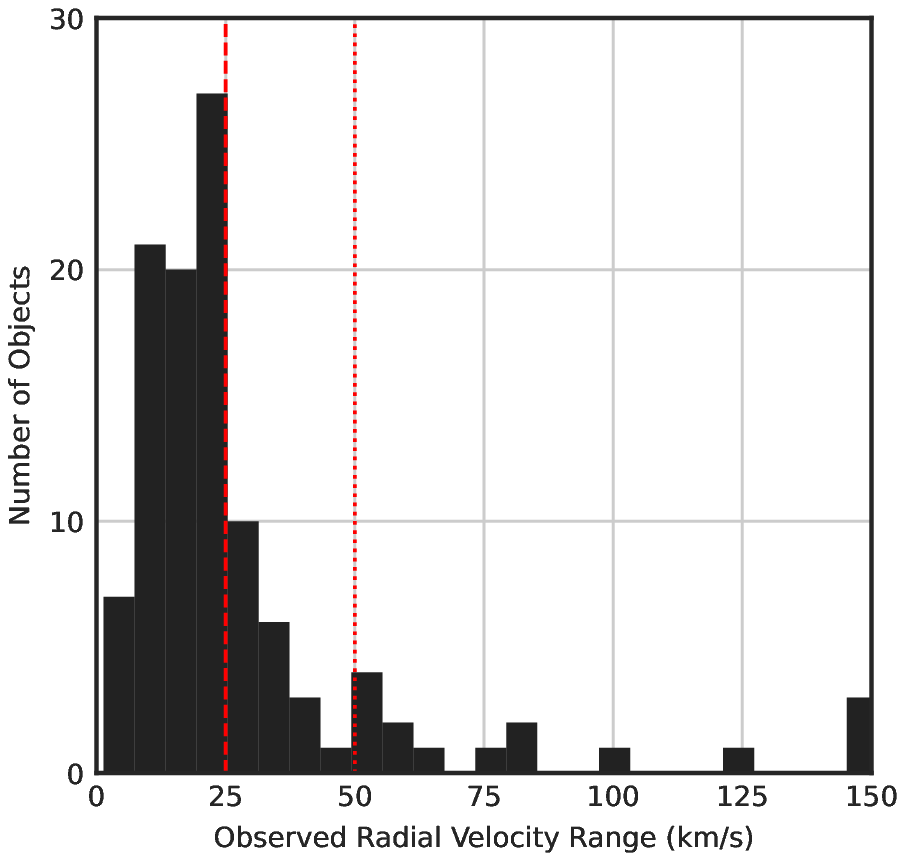}}
    \caption{(\textit{Top Panel}) Number of epochs for each object, (\textit{Bottom Panel}) Distribution of observed RV ranges, with vertical lines marking the $25$\kms and $50$\kms cuts.}
    \label{fig:epochs}
  \end{center}
\end{figure}
\begin{figure*}
  \begin{center}
    \resizebox{\hsize}{!}{\includegraphics{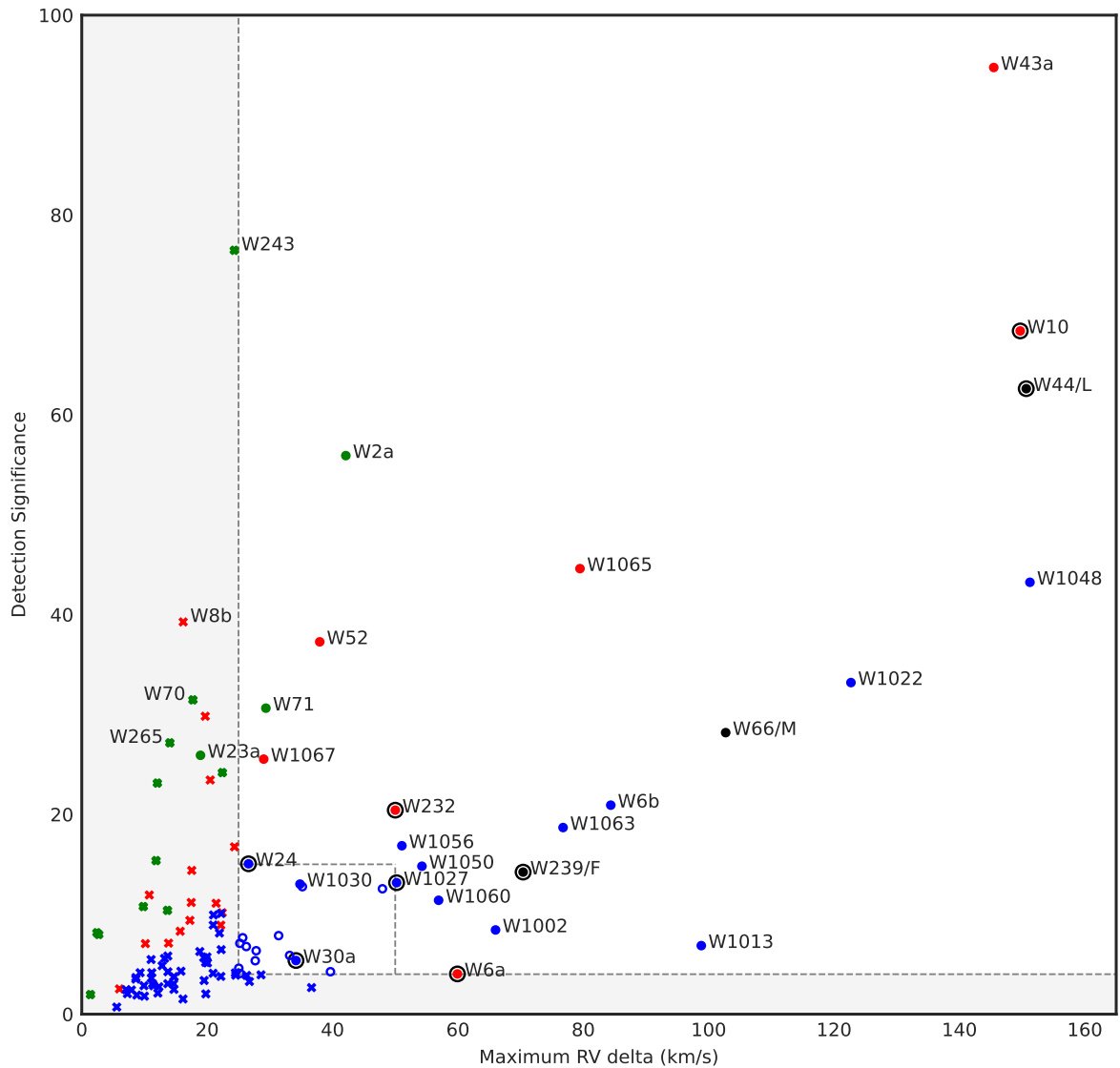}}
    \caption{$\Delta$RV vs. $\sigma_\textrm{det}$. Filled circles represent binaries, open circles represent \textit{candidate} binaries, and crosses represent systems that do not meet the dual criteria described in Section~\ref{sec:cands}. Wolf-Rayet stars are marked in black, O-type stars in blue, B0-B1.5 stars in red, and later types are in green. X-ray detections from Paper~II are marked with a black circle. Dashed lines and shaded areas mark the cutoffs described in the text. Note that \object{W13}, \object{W36}, \object{W53a}, and \object{W1021} have RV ranges that lie outside the bounds of this figure.}
    \label{fig:rv-sigma}
  \end{center}
\end{figure*}

Of the 151 objects listed in Table~\ref{tab:spectype}, 6 are cluster members with only a single
epoch of observation, and 8 are field stars. A 2.26~day eclipsing binary\footnote{Located near the
core of the cluster at J16:47:05.79~-45:51:33.3} reported by \cite{bonanos07} is tentatively
identified as a cluster member based on the strength of the 8620\ang DIB, but cluster membership
cannot be confirmed: this object is extremely faint ($I=16.2$) and cannot be analysed further. In
addition, \object{W9} (sgB[e]; \citealt{clark13}) and \object{W1004} (Oe/Be; Paper~III) show
emission-line spectra devoid of photospheric absorption features suitable for RV measurement, while
21 Wolf-Rayet targets are discussed separately in Section~\ref{sec:wr}. We therefore examine a
population of 113 O--M stars, of which 105 are OB stars: histograms of the number of epochs of
observation and the RV ranges observed are shown in Figure~\ref{fig:epochs}, while full results from
this RV analysis are shown in Figure~\ref{fig:rv-sigma}. 

We find that 34 targets (30\% of the sample of 113 objects) do not display statistically-significant
RV changes\footnote{RV variability is considered statistically-significant if we can reject the null
hypothesis that no RV changes are present at the $99.99\%$ confidence level.}. Almost all are O-type
stars from luminosity class Ib--III, with correspondingly reduced $S$/$N$ and increased fitting
error, and many have limited sampling, implying that we may not be observing the full RV range. Four
objects fail the statistical-significance test despite exceeding the $\Delta\textrm{RV}>25$\kms
threshold. Three are faint, very broad-lined stars with large measurement errors, while the fourth
is an O9.5~Iab star with only three epochs of observation; in each case we also find
$\sigma_{\textrm{det}}<4$. Only one B-type star (\object{W1009}; B0~Ib) does not display
statistically-significant RV variability. 

Of the remaining 79 targets, 43 (or 38\% of the sample) do not meet the dual requirements of
Section~\ref{sec:cands}: in all but one case we find $\sigma_{\textrm{det}}>4$ but
$\Delta\textrm{RV}<25$\kms, while one object meets neither requirement. These targets are
predominantly B0~Ib or later, with the increased luminosity, narrow Paschen series lines, and
apparently-ubiquitous presence of photospheric pulsations in B-type supergiants leading to increased
detection significance ($\sigma_\textrm{det}$), although some O-type stars of luminosity class I or
II also fall into this group. While these objects cannot be classified as spectroscopic binaries
based on their RV variability, we note that one object in this group (\object{W23a}; B2~Ia) displays
a composite spectrum that confirms its binary nature; it is discussed further in
Section~\ref{sec:pulse}.     

Finally, 37 targets (or 32\% of the sample) fully satisfy the criteria for binary classification.
Most interesting are the 23 objects with $\sigma_{\textrm{det}}>15$ or $\Delta\textrm{RV}>50$\kms:
these targets offer high $S$/$N$ or large changes in RV, making them suitable for more detailed
analysis in Section~\ref{sec:individual}. We note that these selection criteria favour the narrow
line widths and high luminosity (and hence high $S$/$N$) of B supergiants, although these selection
criteria are again also met by a number of O9.5~II stars, as well a few targets from earlier
spectral types that have large RV amplitudes. In general, less can be said about the 14 objects that
pass our selection criteria but lie in the range $25<\Delta\textrm{RV}<50$\kms and
$5<\sigma_{\textrm{det}}<15$. Two examples (\object{W30a} and \object{W1030}) have very strong
secondary indicators of binarity and an extensive baseline of observations that permit further
analysis; these two objects are also discussed in Section~\ref{sec:individual}. The remaining 12
objects are all O-type stars, with only one classified as a supergiant, and display low
semi-amplitudes, broad lines, and low $S$/$N$: with fewer than ten epochs of observation, it is
impossible to select between candidate orbital solutions with any confidence. Therefore, while it is
possible to identify statistically-significant RV changes in early-type objects, along with
supporting evidence for binarity based on the $R$- and $I$-band spectral morphology (see also
Paper~III), it is usually not possible to analyse the low-luminosity O9~III population in detail: we
discuss these systems further in Section~\ref{sec:probable}. 

\subsection{Average cluster radial velocity}
We determine the cluster systemic velocity from the error-weighted average of the RVs of 58 OB stars
in our survey that have $\Delta$RV$<25$\kms and more than five epochs of observation, with the
latter criterion ensuring that time-averaging limits any extremes of pulsational variability. The
absorption lines used for RV measurement are expected to form quite close to the photosphere and
therefore no RV offset due to wind-filling is expected, implying that measured RVs should be
consistent across this population. The resulting cluster RV of $-43.1$\kms is in good agreement with
the mean RV of $-44.7\pm1.9$\kms found from interstellar C$_2$ Phillips~(2-0) absorption features in
the spectra of the luminous B2--4~Ia supergiants \object{W57a}, \object{W70}, and \object{W71},
suggesting the absorbing material is likely local to the cluster. It is also in agreement with the
time-averaged \textit{Gaia} EDR2 RVs for the cool hypergiants \object{W4}, \object{W8a},
\object{W26} and \object{W265}, implying no substantial systematic effects are present in our
observations.  Detailed analysis of the cluster RV and its implications for the distance and
location of Wd1 will be presented in Negueruela~et~al.~(in~prep.).

\section{Binaries}
\label{sec:individual}
\begin{table}[!htpb]
  \caption{The 26 targets discussed in Section~\ref{sec:individual}. The three objects listed as
  \textit{Morphology} or \textit{X-Ray} do not meet the criteria defined in Section~\ref{sec:cands},
  but have spectra or X-ray luminosities that unambiguously indicate a binary nature.}
  \label{tab:bins}
  \begin{center}
  \begin{tabular}{lccl}
  Type & Count & New? & Identification\\
  \hline\hline
  &\\
  Eclipsing SB2       & 2  & 0 & W36$^a$, W1021$^a$ \\
  SB2                 & 2  & 1 & W10$^b$, W53a \\
  B0--B2.5~Ia         & 8  & 6 & W2a, W6a, W43a$^c$,\\ 
                      &    &   & W52, W71, W232,\\ 
                      &    &   & W1065$^c$, W1067\\
  O9--9.5~Iab--Ib     & 3  & 3 & W24, W1027, W1048\\
  O9--9.5~II-III      & 6  & 6 & W6b, W1022, W1050,\\ 
                      &    &   & W1056, W1060, W1063\\
  O + O               & 2  & 2 & W1002, W1013\\                    
                      \hline
  \\
  Morphology          & 2  & 2 & W23a, W1030 \\
  X-Ray               & 1  & 0 & W30a$^d$ \\
  \hline
  \\
  Total               & 26 & 20\\
  \end{tabular}
  \end{center}
Previously confirmed as binary in: $^a$\cite{bonanos07}, $^b$\cite{neg10}, $^c$\cite{ritchie11}, $^d$\cite{clark19a}.
\end{table}

Table~\ref{tab:bins} lists the objects considered in this section. Of the 23 OB stars with high
detection significance ($\sigma_\textrm{det}$) and/or large RV amplitudes, \object{W36} and
\object{W1021} are eclipsing SB2s first reported by \cite{bonanos07}, \object{W10} was identified as
a SB2 by \cite{neg10}, and \object{W43a} and \object{W1065} were identified as SB1s by
\cite{ritchie11}. Of the remainder, six are B0--B2.5~Ia stars, while twelve (including one
newly-identified SB2) are O-type stars. In addition, we include three objects that do not meet the
criteria in Section~\ref{sec:cands} but are unambiguously binary based on secondary indicators:
\object{W23a} (B2~Ia) and \object{W1030} (O9.5~Ib) have spectra that are inconsistent with a single
spectral type, while the high X-ray luminosity of the blue straggler \object{W30a} (O4--6~Ia$^+$)
implies it is a colliding-wind binary. 

\subsection{W10}
\object{W10} was noted as a SB2 (B0.5~I + OB) by \cite{neg10}, based on VLT/FORS observations that
showed broad H$\alpha$ emission and double-lined He~I\lamlam6678,7065. It is also detected at X-ray
\citep{clark08} and radio \citep{andrews} wavelengths. We observed \object{W10} on four nights in
2011 and a further four nights in 2013, with the dataset comprising one spectrum in HR16 mode
(showing a single He~I\lam7065 line) and ten spectra in HR21 mode. Although \object{W10} appears
single-lined in most observations, with RVs in the range $-$50~\kms to $-$65~\kms, on MJD~56483.20
the Paschen series lines are clearly double, with the more luminous primary at $+$83$\pm$2~\kms and
a weaker secondary at $-$164$\pm$6~\kms (see also Figure~1 in Paper~II). 

These observations suggest the system has a high eccentricity, with the B0.5~I primary dominating
the composite spectrum in most epochs and the secondary only visible near periastron. Such a
configuration is also required to understand the X-ray properties of the system: \object{W10}
appears softer and less X-ray luminous than other SB2s in Wd1, possibly reflecting observations at a
phase where the wind collision zone was weak or obscured. Unfortunately, any orbital fit is strongly
driven by the single epoch of observation near periastron, and we cannot distinguish between a
number of potential periods in the range $\sim5-20$~days. \object{W10} appears single-lined on
MJD~56494.08, implying a rapid periastron passage, but further observations near periastron are
required to securely determine orbital parameters. However, we can estimate a mass ratio of
$\sim$0.90$\pm$0.05, and we tentatively classify the secondary as $\sim$O9.5~II based on the
relative strength of the Pa-11 lines. The presence of a weak He~I\lam8845 absorption line confirms a
$\sim$B0.5~Ia classification for the primary.  

\subsection{The peculiar B0~Iab star W232}
\label{sec:w232}
\object{W232} (=C07-X6; \citealt{clark08}) was identified as probable binary in
Paper~I\footnote{Note that \object{W232} was incorrectly identified as W234 in Paper~I.}, based on
the presence of significant RV changes on short timescales. It was classified as B0~Iab based on its
$R$-band spectrum \citep{neg10}, although the higher Paschen series lines suggest a slightly earlier
$\sim$O9.5~Ib spectral type may be appropriate. \object{W232} was included on the \textit{bright},
\textit{faint1}, and \textit{faint2} target lists, and we obtained 24 RV measurements in 2008--2009,
along with a further 12 RV measurements in 2013. In a number of cases multiple observations were
obtained on the same night, with low-level line profile variability leading to differences in
measured line centres of up to 5\kms in consecutive spectra: this is unlike other objects that were
observed repeatedly on the same night, such as \object{W71} and \object{W1065}, where consecutive
spectra are in excellent agreement. 

\begin{figure}
  \begin{center}
    \resizebox{\hsize}{!}{\includegraphics{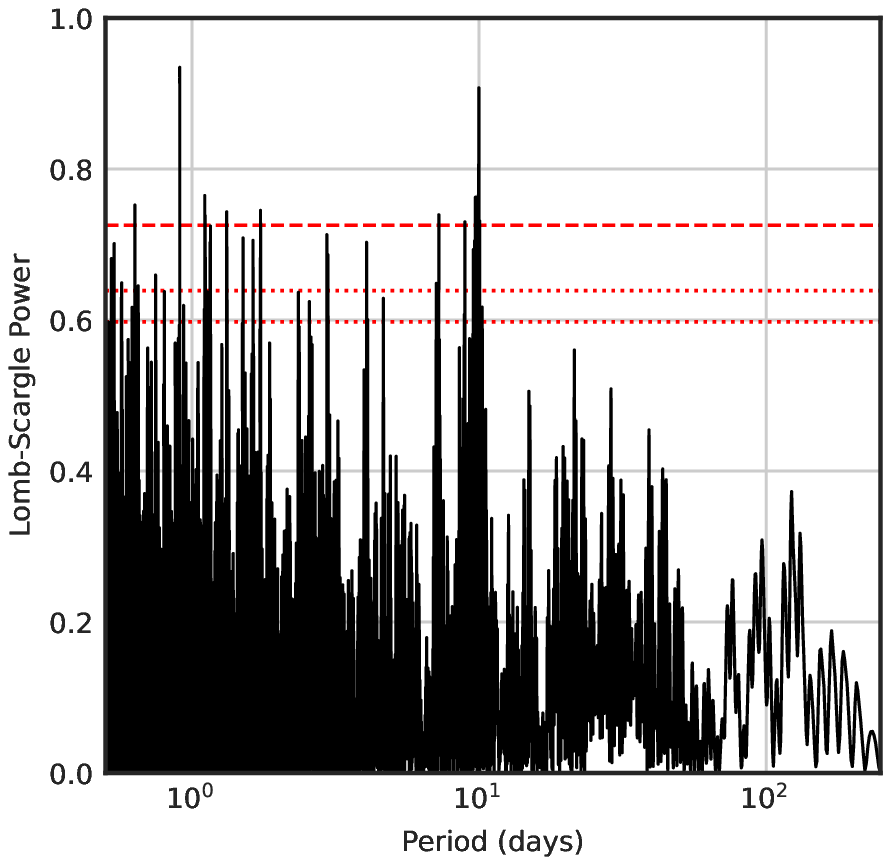}} 
    \resizebox{\hsize}{!}{\includegraphics{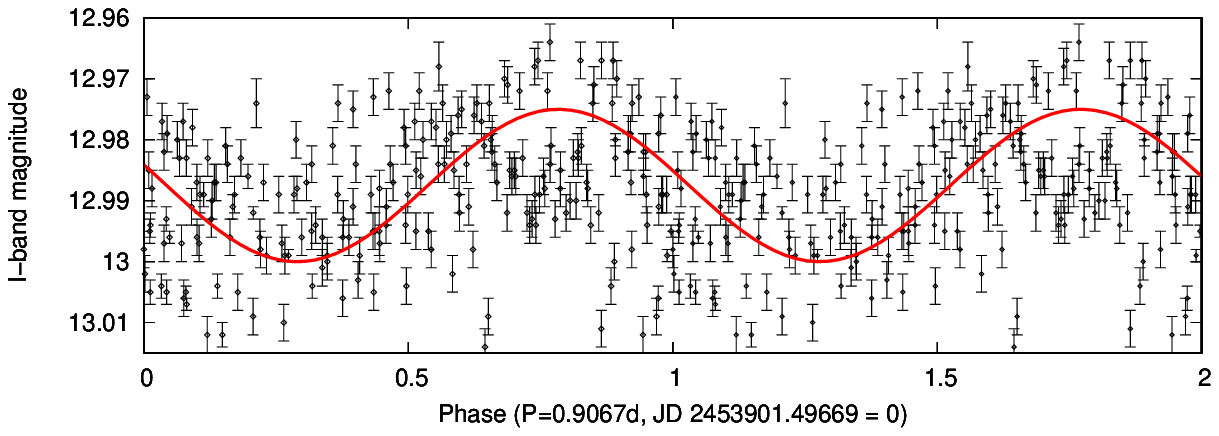}}
  \caption{(\textit{Top}) Lomb-Scargle Periodogram for \object{W232}, showing strong peaks in the RV data at $0.9067$~days and $9.980$~days. (\textit{Bottom}) Photometry from \cite{bonanos07} folded on to the 0.9067~day period. The line marks the best-fit sinusoid.}
  \label{fig:W232}
  \end{center}
\end{figure}

The Lomb-Scargle periodogram shown in Figure~\ref{fig:W232} reveals two strong periodicities at
0.9067~days and 9.980~days, which remain consistent across the 2008--2009 and 2013 datasets. The
former appears inconsistent with an orbital interpretation, with the expected radius of a B0~Iab
primary \citep{martins} implying that any companion would be almost engulfed. However, \object{W232}
appears very similar to other O9.5-B0 supergiants in Wd1, and does not display the complex, blended,
and time-varying spectra seen in very short period interacting systems such as \object{W6a} or
\object{W53a}. Nevertheless, if the photometry of \cite{bonanos07} is folded onto a 0.9067~day
period then a clear low-level sinusoidal variation is apparent (see the lower panel of
Figure~\ref{fig:W232}), suggesting this periodicity is not spurious, and its persistence from
photometry obtained in 2006 to RV data obtained in 2008--2009 and 2013 implies that the underlying
process is stable and therefore unlikely to represent a transient or quasi-periodic photospheric
process. The alternative periodicity provides an orbital solution with $P=9.9778\pm0.0009$~days,
$K_1=14.7\pm2.0$\kms, $\gamma=-45.3\pm1.5$\kms, $e=0.41\pm0.06$, and $\omega=256\pm42$~degrees.
Scatter around the best-fit solution is present, reflecting rapid line-profile variability
superimposed on orbital modulation.

\object{W232} is somewhat more X-ray luminous than expected for an isolated star \citep{clark08},
suggesting a binary interpretation is appropriate, and we note that the sgB[e] binary
\object{GG~Carinae} also displays stable photometric and RV variability on a period of
$\sim$1.53~days that is well below its $\sim$31~day orbital period \citep{porter21}. \object{GG~Car}
has a significantly eccentric orbit, and these short-period pulsations result from tidally-driven
oscillations at the $l=2$ $f$-mode frequency, with the largest variability occurring at periastron.
In the case of \object{W232}, we note that there is an exact integer ratio between the two
identified periods, suggesting the 0.9067~day period is a tidally excited oscillation in a $g$-mode
of the primary that arises from a resonance between an orbital harmonic and a mode frequency
\citep{willems02}. This tidal resonance would explain both the long-term stability of the
$0.9067$~day oscillation and the rapid line profile variability, which may arise from higher order
$g$-modes. The combination of significant eccentricity, short period, and tidally-excited
oscillations suggest that \object{W232} may therefore be an example of a massive `heartbeat star'
\citep{fuller17}; if confirmed, it would represent one of the most massive examples known.

\subsection{Early-B supergiant SB1s: W43a, W52, W1065, and W1067}
\label{sec:earlyB}
\begin{figure}
\begin{center}
\resizebox{\hsize}{!}{\includegraphics{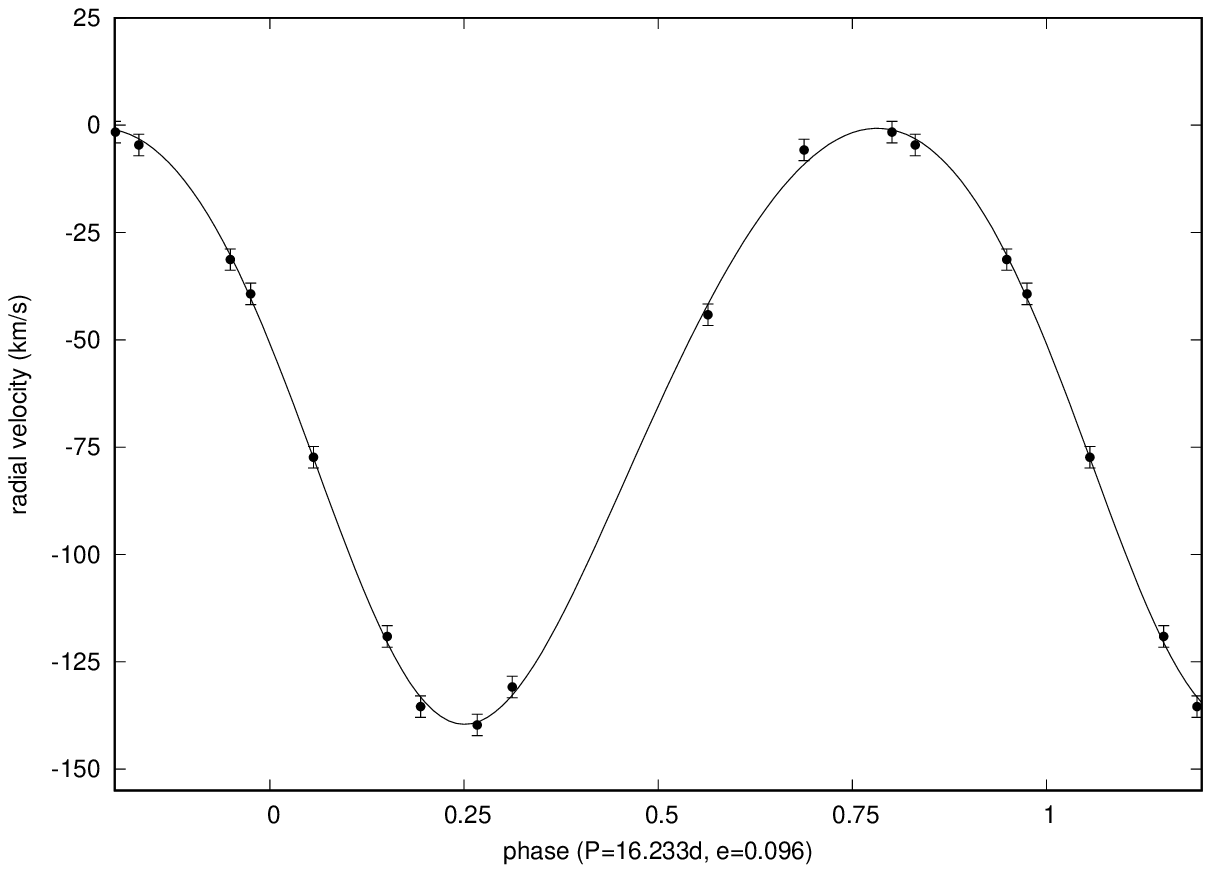}}
\resizebox{\hsize}{!}{\includegraphics{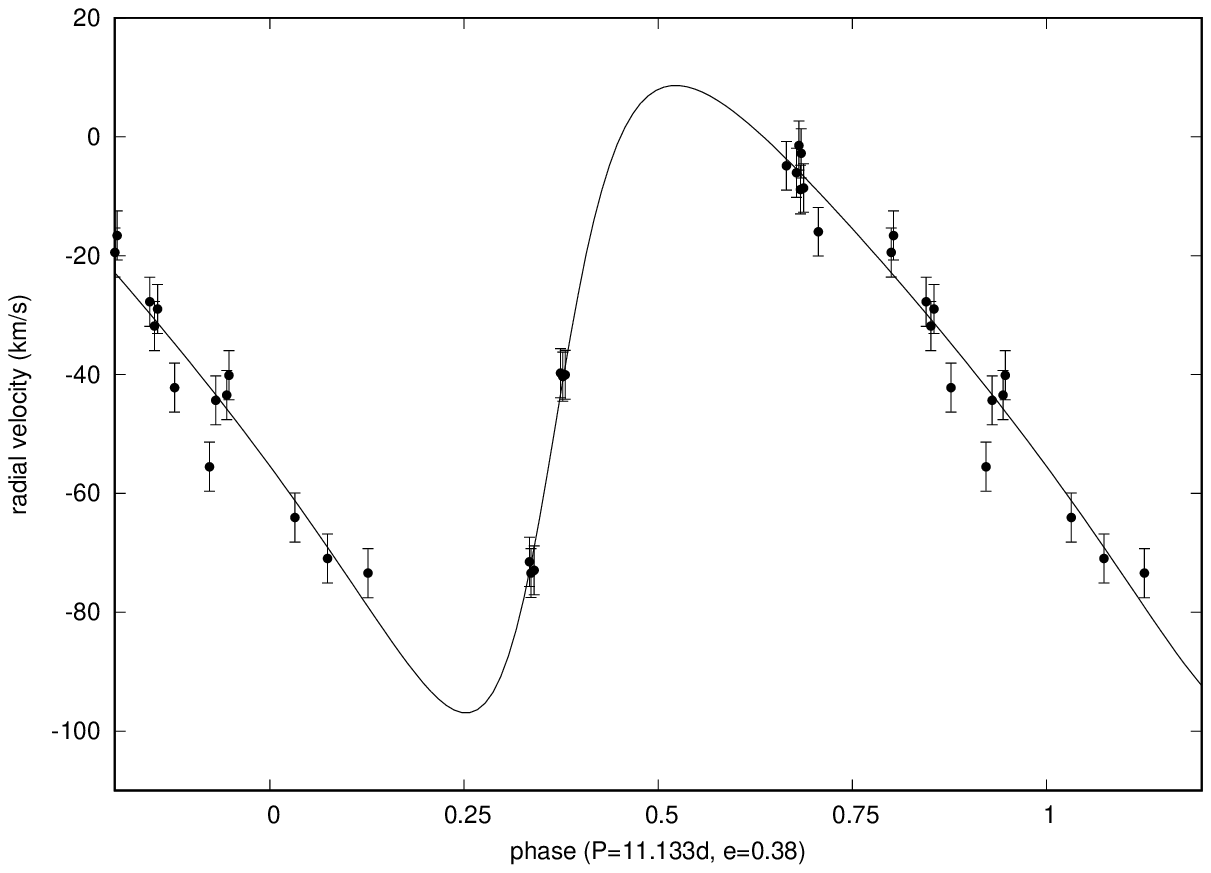}}
\caption{RV curve for the B0 supergiants \object{W43a} (\textit{Top Panel}) and \object{W1065} (\textit{Bottom Panel}).}
\label{fig:earlyB}
\end{center}
\end{figure}
\object{W43a} is a B0~Ia supergiant that was identified as an SB1 displaying low-level pulsational
variability in Paper~I (see also \citealt{ritchie11}). We find a period $16.2333\pm0.0048$~days,
with $K_1=69.4\pm0.9$~\kms and $\gamma=-64.4\pm0.8$\kms, while the orbit appears slightly eccentric,
with $e=0.096\pm0.017$ and $\omega=149\pm9$~degrees. The RV curve is plotted in
Figure~\ref{fig:earlyB}. No eclipses are seen in the photometry of \cite{bonanos07}, but complete
phase coverage is not available and an eclipsing system cannot be ruled out. 

\object{W1065} (B0~Ib; previously identified as \object{W3003}) was also identified as a
short-period binary in Paper~I. It was included in the 2008--2009 \textit{bright}, \textit{faint2}
and \textit{faint3} fields, giving a total of 23 observations, while four additional observations
were obtained in 2013. The orbital period is $11.1330\pm0.0012$~days, with $e=0.38\pm0.05$,
$\omega=256\pm13$~degrees, $\gamma=-39.4\pm2.6$~\kms, and $K_1=52.8\pm3.6$~\kms. The semi-amplitude
implies a lower inclination and/or lower-mass secondary than \object{W43a}. The RV curve is again
plotted in Figure~\ref{fig:earlyB}. \object{W1067} (B0~Ib; previously identified as \object{W3004})
appears very similar to \object{W1065}. It was duplicated on the \textit{bright} and \textit{faint2}
lists, providing 17 observations in 2008--2009, and we find a $6.125\pm0.021$~day period with
$K_1=13.2\pm3.8$\kms and $\gamma=-47.0\pm0.9$\kms. As with \object{W1065}, the best-fit solution is
eccentric, with $e=0.52\pm0.16$ and $\omega=355\pm15$~degrees. Two low-resolution $R$-band spectra
from 2013 are available, and we note H$\alpha$ in emission with a weak P~Cygni profile superimposed
on a broader base, which is inconsistent with an isolated B0~Ib star \citep{neg10}. He~I absorption
features appear single lined.  

Finally, \object{W52} (B1.5~Ia) was only observed on five nights in 2013, but RVs obtained on
consecutive nights (MJD=56511.2 and 56512.2) confirm that \object{W52} is a short-period binary.
There is insufficient data for a satisfactory period search, but Bonanos (2007) finds \object{W52}
to be a 6.7~day periodic variable, and we tentatively identify a $6.624\pm0.108$~day period, with a
semi-amplitude of $\sim$25~\kms and an apparently-circular orbit. The RV curve is not well sampled,
but the close correspondence between the tentative RV solution and the photometric modulation
suggests that the correct orbital period may have been identified. Additional observations are
required to determine a robust solution. 

None of these four systems are detected at X-ray wavelengths \citep{clark08}, and none display any
spectroscopic features attributable to a secondary, which are likely OB~III-V objects with a lower
luminosity and weaker wind than their early-B supergiant primaries.  

\subsection{W1048}
\begin{figure}
  \begin{center}
  \resizebox{\hsize}{!}{\includegraphics{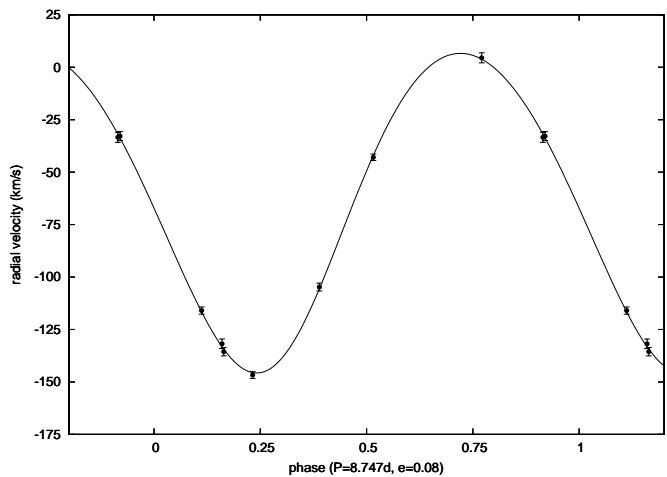}}
  \caption{RV curve for the O9.5~Ib supergiant \object{W1048}.}
  \label{fig:W1048}
  \end{center}
\end{figure}
\object{W1048} was identified as a periodic variable by \cite{bonanos07}. Nine observations were
obtained with FLAMES in 2013, including observations on consecutive nights, and the system displays
large RV changes that confirm its binary nature. It was not possible to reconcile the RV data with a
5.2~day periodicity reported by \cite{bonanos07}, and we instead find an orbital period of
$8.747\pm0.021$~days with $K_1=76.2\pm1.0$\kms and $\gamma=-64.0\pm0.6$\kms. Like \object{W43a}, the
system appears slightly eccentric, with $e=0.08\pm0.03$ and $\gamma=205\pm8$~degrees. The RV curve
is shown in Figure~\ref{fig:W1048}. Paper~III classifies \object{W1048} as B1.5~Ia, but this appears
erroneous: \object{W1048} displays a slightly earlier spectrum than \object{W1065} (B0~Ib) and is
reclassified as O9.5~Ib accordingly. 

Like the early-B supergiants discussed in Section~\ref{sec:earlyB}, \object{W1048} is not detected
at X-ray wavelengths (Paper~II), but the large semi-amplitude implies a massive secondary at
favourable inclination. No archival observations are available, precluding a broader search for
features attributable to the secondary, but refolding the $I$-band photometric observations of
\cite{bonanos07} onto our $8.747$~day period suggests that a shallow $\sim$0.1-magnitude eclipse may
be present, with a near-contact configuration. The photometric data have considerable scatter and
are not suitable for detailed modelling, but a further programme of spectroscopic and photometric
observations of \object{W1048} would place constraints on the mass of objects entering the
supergiant phase.

\subsection{W1030}
\begin{figure}
  \begin{center}
    \resizebox{\hsize}{!}{\includegraphics{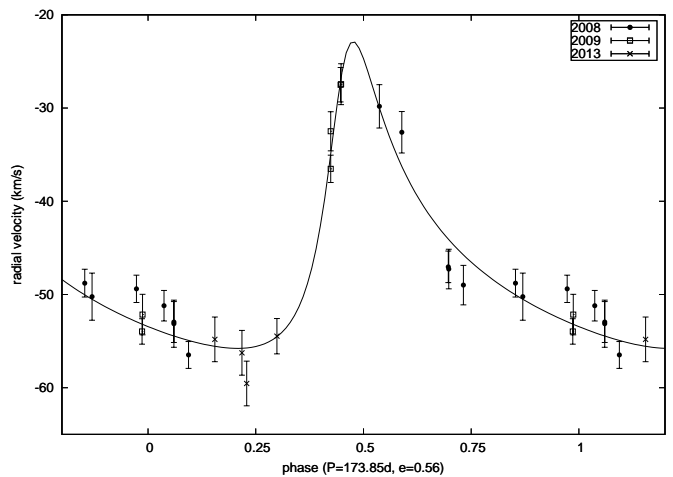}}
    \caption{RV curve for the 174~day O9.5~Ib binary \object{W1030}. Observations from 2008 (filled circles), 2009 (open squares), and 2013 ($\times$ symbols) are indicated.}
    \label{fig:w1030}
  \end{center}
\end{figure}
\object{W1030} (O9.5~Ib; previously named \object{W3005} in Paper~I) is located $\sim2'$ south of
the cluster core, corresponding to a projected distance $\sim$3pc at a distance of 5kpc. It was
duplicated on the 2008--2009 \textit{bright} and \textit{faint1} lists and was observed 18 times in
that period. Four additional spectra were obtained in 2013. $R$-band spectra show infilled
H$\alpha$, while in the $I$-band C~III~\lam8500 appears unexpectedly weak for an O9.5 supergiant and
it also appears too luminous for its O9.5~Ib classification. \object{W1030} may be a $\sim$B0~Ia
star with a spectrum diluted by a $\sim$O9.5~II companion, although no direct evidence for the
secondary is observed.  

RVs obtained in 2008 show a steady increase from $-30$\kms to $-56$\kms, with lack of any
significant excursions on timescales of days--weeks precluding a short orbital period. We find an
orbital period of $173.85\pm0.72$~days, by far the longest period currently identified in Wd1, with
$K_1=16.5\pm1.4$\kms and $\gamma=-46.6\pm0.6$\kms. The orbit is eccentric, with $e=0.56\pm0.05$ and
$\omega=322\pm7$~degrees, and the periastron passage is fortuitously well sampled by the  2008 and
2009 observations.  

\subsection{O9.5~II SB1s: W1022, W1050, W1056, and W1060}
\label{sec:brightgiants}
\object{W1022}, \object{W1050}, \object{W1056}, and \object{W1060} are four very similar systems,
with spectra close to W86 (O9.5~Ib) but with broader wings to the Paschen series lines and somewhat
lower $I$-band magnitudes, leading to an O9.5~II classification. Low-resolution FORS2/MXU spectra
are available for \object{W1022} and \object{W1050}, with both objects appearing single-lined with
H$\alpha$ in absorption (see Figure~6 in Paper~III). A low-resolution LR6-mode FLAMES spectrum of
\object{W1056} shows narrow H$\alpha$ and [N~II] nebular emission lines superimposed on a broad,
possibly partially-infilled, photospheric H$\alpha$ absorption line, and we note that \object{W1056}
is detected in ATCA observations by \cite{andrews}. No additional spectroscopy exists for
\object{W1060}. All four objects were observed in the 2008--2009 and 2013 campaigns, providing a
five-year baseline with either ten or eleven observations. All display $\Delta\textrm{RV}>50$\kms,
confirming them as binaries, although period determination is harder: only one object has a
\cite{baluev} false alarm probability below $1\%$, with the other three objects only having
\textit{candidate} periods. None are reported as X-ray detections in Paper~II.    

\object{W1022} shows some of the largest RV changes in our sample, with $\Delta\textrm{RV}>120$\kms
and a delta of more than 50\kms in the space of two nights that confirms a short-period system.
Somewhat surprisingly, the orbital solution is not very well constrained. We find a candidate period
of $6.175\pm0.004$~days with \cite{baluev} false alarm probability $\sim1.2$\% but cannot fully
exclude other potential periods in the 4--9~day range, although all have false alarm probabilities
$\gsim10\%$. The semi-amplitude $K_1=73.2\pm18.4$\kms and $\gamma=-36.6\pm9.8$\kms are also
uncertain due to limited sampling around phase $\sim$0.25. A near-circular orbit is favoured,
although other solutions with $e\le0.3$ cannot be ruled out. Further observations are required to
fully determine the parameters of this system.

\object{W1050} has a candidate period of $4.063\pm0.003$~days, with $K_1=27.6\pm3.0$\kms, and
$\gamma=-51.0\pm1.3$\kms. The orbit is eccentric, with $e=0.56\pm0.10$ and $\omega=161\pm9$~degrees.
We find several potential periods for \object{W1056}: the strongest candidate period is at
$53.134\pm0.031$~days, with $K_1=28.7\pm3.1$\kms, $\gamma=-41.8\pm1.1$\kms, $e=0.31\pm0.09$, and
$\omega=55\pm7$~degrees, although an alternative period of $\sim9.97$~days cannot be ruled out. This
shorter period leads to very similar values of $K_1$, $\gamma$, and $e$, although $\omega\sim0$ is
required. We provisionally adopt the longer period, pending further observations. Finally, we find a
period of $3.238\pm0.003$~days for \object{W1060}, with false alarm probability $<1\%$. The orbit is
near-circular, with $e\lsim0.1$, $K_1=28.7\pm6.1$\kms, and $\gamma=-59.3\pm2.6$\kms. The low
semi-amplitude of these three objects suggests unfavourable inclinations and/or a low-mass
companions.

\subsection{O9~III binaries: W6b and W1063}
\label{sec:O9giants}
\begin{figure}
  \begin{center}
  \resizebox{\hsize}{!}{\includegraphics{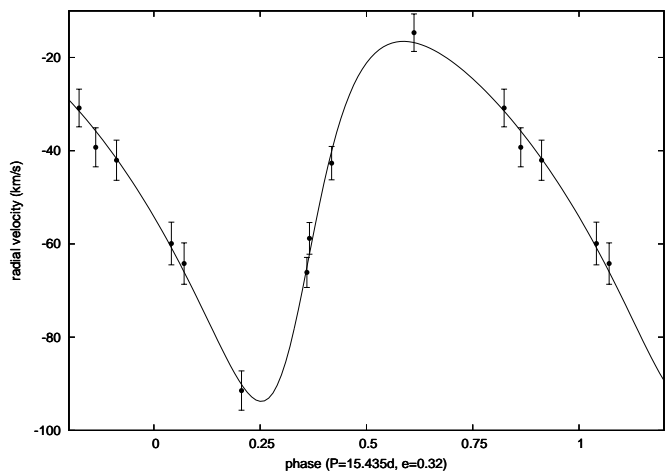}}
  \caption{RV curve for the O9.5~III star \object{W1063}.}
  \label{fig:O95III}
  \end{center}
  \end{figure}
\object{W6b} was observed on six nights in 2008--2009 and on a further three nights in 2013.
Although it is classified as O9~III and is one of the faintest objects in our sample ($I=15.25$;
\citealt{cncg05}), the Pa-11 line is well defined and we find a candidate $2.653\pm0.002$~day
period, although other potential periods with $P<10$~days exist. The best-fit solution has
$K_1=41.6\pm4.0$\kms, $\gamma=-64.2\pm2.4$\kms, $e=0.29\pm0.06$, and $\omega=249\pm21$~degrees, with
the offset systemic velocity suggesting we may have incomplete sampling of the full range of RVs. We
caution that the solution should be considered provisional at best. There is no indication of a
secondary, which we assume must still be on the MS. 

\object{W1063} displays a very similar $I$-band spectrum to \object{W6b}, and receives the same
O9~III classification. It was observed on six nights in 2008--2009 and on a further four nights in
2013 and we find a candidate period of $15.435\pm0.073$~days with $K_1=38.6\pm2.1$\kms,
$\gamma=-47.8\pm1.0$\kms, $e=0.32\pm0.05$, and $\omega=233\pm7$~degrees. The false alarm probability
is $\sim4\%$, although no alternative periods are identified. The wings of the Pa-11 line appear
asymmetric at maximum separation, suggesting a possible contribution from a an OB~IV--V companion.
Two low-resolution $R$-band spectra obtained in 2013 show H$\alpha$ and HeI\lamlam6678,7065 in
absorption, with RV changes also apparent in the He~I lines. 

\subsection{B2--2.5 supergiants: W2a, W23a, and W71}
\label{sec:pulse}
\begin{figure}
  \begin{center}
    \resizebox{\hsize}{!}{\includegraphics{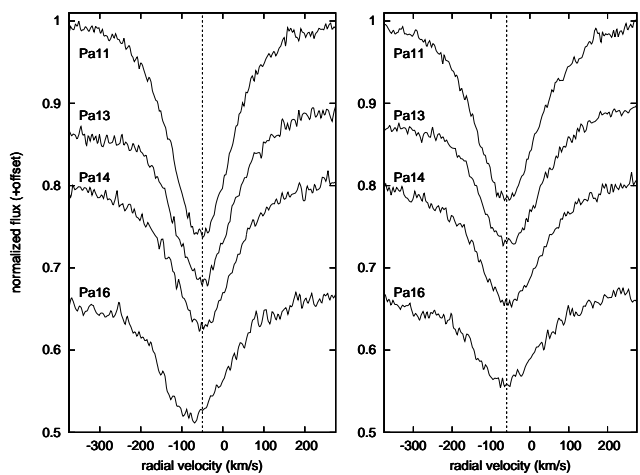}}
    \caption{Montage of Paschen series absorption lines in \object{W23a} (\textit{left panel}) and \object{W2a} (\textit{right panel}). Vertical lines mark the average of the Pa-11 and Pa-13\dots15 lines (Pa-12 is omitted due to the influence of the C$_2$ Phillips (2-0) system). Note the discrepant line centre and greater depth for Pa-16 in \object{W23a} due to blending with C~III~\lam8500; this is not seen in \object{W2a} or other luminous B supergiants, and implies the existence of a O9~I secondary.}
    \label{fig:w23a}
  \end{center}
\end{figure}
\object{W23a} was proposed as a candidate binary in Paper~I, and was noted by \cite{neg10} as
displaying \textit{features incompatible with a single spectral type \dots the spectrum may include
an earlier type companion}. Our data support this assertion. \object{W23a} displays Pa-15/Pa-16$<1$,
a ratio that is otherwise only seen in objects earlier than B0~I (see Figure~2 in Paper~III), but
strong Pa-14, Pa-15, and He~I\lamlam8845,8777,8583 absorption lines \textit{preclude} a
classification earlier than B2~Ia. In addition, RVs derived from the Pa-11 and Pa-13--15 lines are
in excellent agreement, with individual line centres consistent to within 1\kms, but the Pa-16 line
centre is displaced by $\sim30$\kms (see Figure~\ref{fig:w23a}), with this offset and the anomalous
Pa-15/Pa-16 ratio indicating blending with a strong C~III\lam8500 absorption line. The strength of
this C~III feature implies an $\sim$O9~I companion \citep{neg10}; lower luminosity classes would not
be apparent in the spectrum of a highly-luminous B2~Ia primary. We note that \cite{cncg05} highlight
variable H$\alpha$ emission from \object{W23a}, although no follow-up $R$-band spectra are
available. If periods less than ten days are excluded as incompatible with the physical sizes of two
\textit{non-interacting} supergiants, then the string-length periodogram finds a possible period of
$\sim$40.5~days. We find a potential orbital solution with $K_1=9.2\pm2.0$ and $e\sim0.2$, although
a circular orbit is not excluded. 

\begin{figure}
  \begin{center}
    \resizebox{\hsize}{!}{\includegraphics{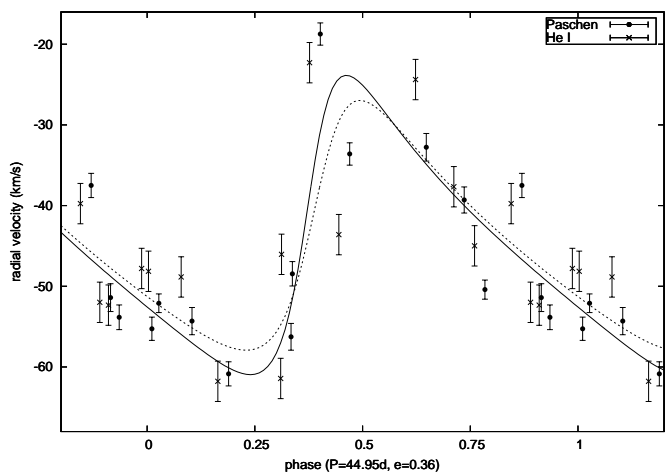}}
    \caption{Candidate RV curves for \object{W2a}. The solid line represents a fit to the Paschen series lines, while the dashed line represents an \textit{independent} fit to He~I\lamlam8583,8845. Scatter around the fit appears to be pulsational in origin.}
    \label{fig:w2a}
  \end{center}
\end{figure}

Both \object{W2a} and \object{W71} satisfy the $\Delta\textrm{RV}>25$\kms cut introduced in
Section~\ref{sec:cands}, and \object{W2a} was proposed as a candidate binary in Paper~I. It is not
possible to fit either object with an unambiguous orbital solution, with significant residuals
remaining for any choice of orbital parameters. Nevertheless, we measure a substantially higher RV
amplitude in these objects than is observed in any of the other putative pulsators of similar
spectral type (e.g. \object{W8b} or \object{W57a}), and these objects appear to display
\textit{both} pulsational variability and orbital motion. We tentatively identify a $44.95$~day
period for \object{W2a}, based on observations from both the 2008--2009 and 2013 campaigns, noting
that this period can be independently determined from both the Paschen series and from strong
He~I\lamlam8583,8845 absorption lines. A moderately eccentric solution ($e\sim0.3$) is preferred.
Substantial scatter is present around the best-fit solution (see Figure~\ref{fig:w2a}), including
variability on timescales of a few days that is inconsistent with a $\sim45$~day period, supporting
the presence of both photospheric and orbital variations. Finally, we find a tentative $36.6$~day
solution with $e\sim0.1$ for \object{W71} based off 18 epochs of data obtained during 2008--2009,
with scatter again present. Neither object displays the anomalous C~III\lam8500 absorption seen in
\object{W23a}, implying secondaries in an earlier, less-luminous state.  

The combination of low semi-amplitude and pulsational variability means that the candidate solutions
for these three objects require additional observations for confirmation. However, with periods in
the $\sim$35--45 day range, all three objects appear potential evolutionary precursors to the
interacting 53.95~day WN9h: binary \object{W44}/\object{WR~L}, which is discussed further in
Section~\ref{sec:wr}.

\subsection{X-ray sources: W24 and W1027}
\label{sec:hard}
\object{W24} (O9~Iab) is a X-ray source that is directly comparable in hardness and luminosity to
the WNVL+OB binary \object{W13} \citep{clark08}. The C~III~\lam8500/Pa-16 blend is variable (see
Figure~4 in \citealt{clark10}), suggesting that C~III~\lam8500 may be augmented by an O-type
secondary in a short-period system. \object{W24} was observed on seven nights in 2008--2009 and a
further five nights in 2013, and we find a candidate period of $6.595\pm0.012$~days with false alarm
probability $0.46\%$, although we cannot fully exclude an alternative solution with a period of
$4.50$~days and false-alarm probability $4.4\%$. The orbital solution has $K_1=13.6\pm0.6$\kms and
$\gamma=-50.4\pm0.4$\kms: given the short period and the massive $\sim$O9 secondary suggested by the
C~III~\lam8500 variability, this implies an almost pole-on inclination. The best fit solution has
$e=0.25\pm0.04$ and $\omega=212\pm8$~degrees, implying a pre-interaction state. 

\object{W1027} (O9.5~Iab) has very similar hardness and intensity to \object{W1040} (=C07-X3;
Paper~II) and \object{W6a}. It was only observed on four nights in 2013, displaying
$\Delta$RV$>50$\kms and change of almost 35\kms in the space of two nights (MJDs 56494.1 and
56496.1) that confirms it as a short-period binary. Variable C~III may also be present in this
object, again suggesting the presence of an O9~I-II secondary, but the limited set of observations
precludes determination of orbital parameters. 

\subsection{Interacting binaries: W6a, W30a, W36, and W53a}
\label{sec:interacting}

\object{W36} (OB+OB; \citealt{neg10}) is a known double-lined eclipsing system ($P=3.18$~days;
\citealt{bonanos07}, \citealt{kb12}), while \object{W6a}, \object{W30a}, and \object{W53a} have very
strong secondary indicators of binarity, including complex, variable H$\alpha$ profiles and/or
strongly enhanced X-ray emission. \object{W30a} (O4--6~Ia$^+$) was identified as a blue straggler
by \cite{clark19a}.  

\object{W6a} was classified as B0.5~Iab by \cite{neg10}, and has previously been identified as a
candidate binary from a 2.20~day modulation in the light curve \citep{bonanos07}, variable H$\alpha$
emission, and moderate X-ray emission \citep{clark08}. It was observed on six nights in 2008--2009,
revealing very weak and broad Paschen series lines. Large uncertainties in fitting line profiles
mean that the requirement that $\sigma_{\textrm{det}}>4$ is only just met, but RV changes with
semi-amplitude $\sim$25$\pm5$\kms confirm the binary nature of the system. No definitive period
determination is possible, but we note that our minimum string length occurs at 2.217~days,
consistent with the 2.20~day modulation of the light curve. Therefore, while orbital parameters
cannot be confirmed, a synthesis of the RV, photometric, and X-ray data strongly suggest a very
short period interacting system.

\object{W30a} was observed on eleven nights in 2008--2009, and on a further five nights in 2013. It
displays $\Delta\textrm{RV}>30$\kms, and although an orbital period cannot be determined with
confidence, we note that a $\sim6.26$~day periodicity is the strongest feature in both the
Lomb-Scargle periodogram and the string-length search, and  suggest this may be the orbital period
of the system: orbital periods above ten days appear excluded (see also discussion in
\citealt{clark19a}). The semi-amplitude is very low, with $K_1=13.2\pm2$\kms, suggesting a near
pole-on viewing angle or very unequal mass ratio; given the blue straggler state and high
spectroscopic mass of the primary, substantial mass transfer may have occurred, leaving a low-mass,
stripped secondary.  

\begin{figure}
  \begin{center}
  \resizebox{\hsize}{!}{\includegraphics{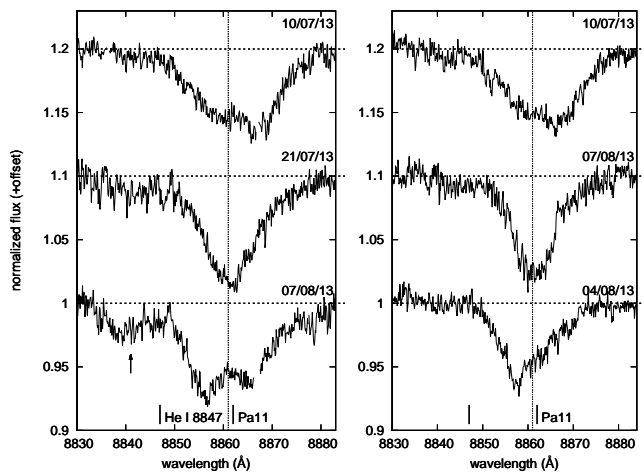}}
  \caption{Montage of the Pa-11 line in \object{W36} ($P=3.18$~days, \textit{Left Panel}) and \object{W53a} (\textit{Right Panel}). Horizontal lines represent the continuum level at each epoch, and the vertical line represents the mean Pa-11 RV for stars in Westerlund~1. Spectra have been rebinned and filtered to reduce noise, and gaps indicate sky line removal as described in Section~\ref{sec:sky}.}
  \label{fig:pa11}
  \end{center}
\end{figure}

\object{W36} was observed on five nights in 2013, fortuitously capturing the system close to both
extremes of the orbit (see Figure~\ref{fig:pa11}). No archival $R$-band spectra are available. At
the midpoint it appears single-lined, although the weak features appear anomalously early for its
luminosity and OB~Ia spectral type (\citealt{neg10} and Paper~III). At quadrature it is clearly
double-lined, although still heavily blended. We estimate semi-amplitudes of $K_1=180\pm20$\kms and
$K_2=220\pm15$\kms, consistent with the results of \cite{kb12}. The deeper-lined component shows
greater RV changes, and we therefore associate this feature with the secondary. He~I\lamlam8733,8845
absorption lines appear to be moving in phase with this component. The contact configuration
\citep{bonanos07} and surprisingly low dynamical masses for its spectral type and cluster context
($16+11$\Msun; \citealt{kb12}) suggest substantial mass loss has occurred in the past\footnote{We
note that \object{W36} is not detected by \cite{fenech} or \cite{andrews}, indicating current
mass-loss rates are low.}. The unusual strength of He~I is consistent with a secondary that has lost
much of its Hydrogen envelope, and \object{W36} appears similar to the 2.185~day overcontact binary
\object{LSS~3074} (O4f$^+$+O6-7:(f):, \citealt{raucq}), which has evolved from a star with initial
mass $30-35$\Msun via non-conservative Roche-lobe overflow, and is now observed in the slow-phase of
Case~B evolution en route to a WR+O configuration. A similar scenario for \object{W36} would be
fully consistent with the cluster environment, and would suggest \object{W36} is an evolutionary
precursor to the WN7+O binaries \object{WR~A} and \object{WR~B}. Further observations aimed at
firmly establishing the current state and evolutionary history of \object{W36} would be valuable. 

\cite{bonanos07} report \object{W53a} to be a semiregular variable with a 1.3~day periodicity in the
light curve, while \cite{clark08} show it to be a strong X-ray source with hardness and luminosity
directly comparable to \object{W36}. Observations were obtained on seven nights in 2013, including
observations on consecutive nights on two occasions. Spectra show the width and profile of the
Paschen series lines to vary considerably, suggesting a SB2 binary with a heavily-blended spectrum
(see Figure~\ref{fig:pa11}) and Pa-11 line profiles vary from night to night, consistent the very
short period implied by the photometric modulation. At its broadest, the spectrum appears to
displays two distinct minima, while at its narrowest the line centre is well defined and lies at
$-40\pm8$~\kms, close to the average systemic velocity for the cluster. We estimate
$K_1=125\pm30$\kms and $K_2=140\pm20$\kms, which would imply a mass ratio $\sim0.9$, noting large
uncertainties in both line centres and that our limited set of observations may not fully sample the
full range of RVs. If confirmed, this would support the presence of an $\sim$O9~I primary and
$\sim$O9~II secondary, as suggested in Paper~II. Although we cannot obtain an orbital solution from
the heavily-blended $I$-band spectra, our results demonstrate the system is a double-lined binary
and are consistent with the 1.3~day periodicity determined from the light curve. The very short
orbital period precludes components later than $\sim$O9~I (cf. \citealt{martins}), and the system
must be in a near-contact configuration. A single low-resolution FORS spectrum is available, showing
broad and weak H$\alpha$ emission and an asymmetric He~I~$\lambda7065$ absorption line. Further
observations in the $R$~band may permit a complete orbital solution to be determined.
 
\subsection{Broad-lined systems: W1002, W1013, and W1021}
\begin{figure}
  \begin{center}
    \resizebox{\hsize}{!}{\includegraphics{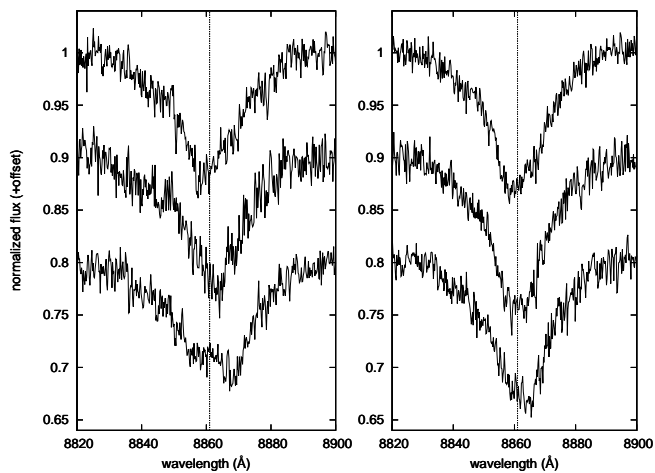}}
    \caption{Montage of the Pa-11 line for \object{W1021} (=B07-DEB1, $P=4.45$~days, \textit{left panel} and \object{W1002} (O9--9.5~II~+~O?; \textit{right panel}). Vertical lines mark the mean Pa-11 RV for stars in Westerlund~1. Spectra are from MJD 56483.2 (\textit{top}), 56494.1, and 56508.1 (\textit{bottom}), and have been rebinned and filtered to reduce noise.}
  \label{fig:deb}
  \end{center}
  \end{figure}

\object{W1021} is a 4.45~day detached, eclipsing binary first reported by \cite{bonanos07}. Like
\object{W53a}, it was observed on seven nights in 2013, displaying an largely-featureless $I$-band
spectrum with very broad Pa-11 and Pa-12. Although the semi-amplitudes of the two components are
directly comparable to \object{W36} \citep{kb12}, the Pa-11 lines can barely be separated, and most
epochs only display an asymmetric, time-varying line profile reflecting two early-type objects.
Without the observations of \cite{bonanos07} and \cite{kb12}, we could do little more than classify
\object{W2021} as a candidate O9~III-V binary based on its $I$-band morphology, with a suggestion
that one component has a semi-amplitude $\ge150$\kms derived from variations in the line profile.
However, \object{W2021} is important in two respects. Firstly, association of this broad-lined and
time-varying $I$-band morphology with a confirmed early-type binary system provides support for our
conjecture that many, if not all, of the systems with similar time-varying morphologies are also
$\sim$O9~III binaries (see Section~\ref{sec:probable} and discussion Paper~III). Secondly, although
the Pa-11 blend is unsuitable for detailed analysis, the orbital solution provided by \cite{kb12}
demonstrates that follow-up observations centred on He~I~\lam7065 would support determination of
orbital solutions for these objects, provided that inclinations are favourable. 

\object{W1002} also displays broad lines with significant changes in line profile (see
Figure~\ref{fig:deb}). The $I$-band luminosity is similar to the O9.5~II systems discussed in
Section~\ref{sec:brightgiants}, but broad, dilute, and asymmetric Paschen series lines are virtually
identical to \object{W1021}, although the greater depth in some epochs suggests the primary has a
slightly later spectral type and this system may contain an O9~II-III primary and an O-type
secondary. The Paschen series lines of the two components cannot be clearly separated, but two
$R$-band spectra are also available, with He~I\lam7065 appearing partially separated in one epoch:
at least one component has a semi-amplitude $\gsim$100\kms. We are unable to determine a potential
orbital period from the available data. 

Finally, \object{W1013} is a very faint target with a similar spectrum to \object{W1002}, with
broad, dilute Paschen series lines leading to an O+O? classification in Paper~III. RV changes of
almost $\sim$100\kms appear to be present, accompanied by changes in line profile, but limited
sampling and very low $S$/$N$ preclude further analysis.  

\begin{figure*}[htp]
  \begin{center}
  \resizebox{\hsize}{!}{\includegraphics{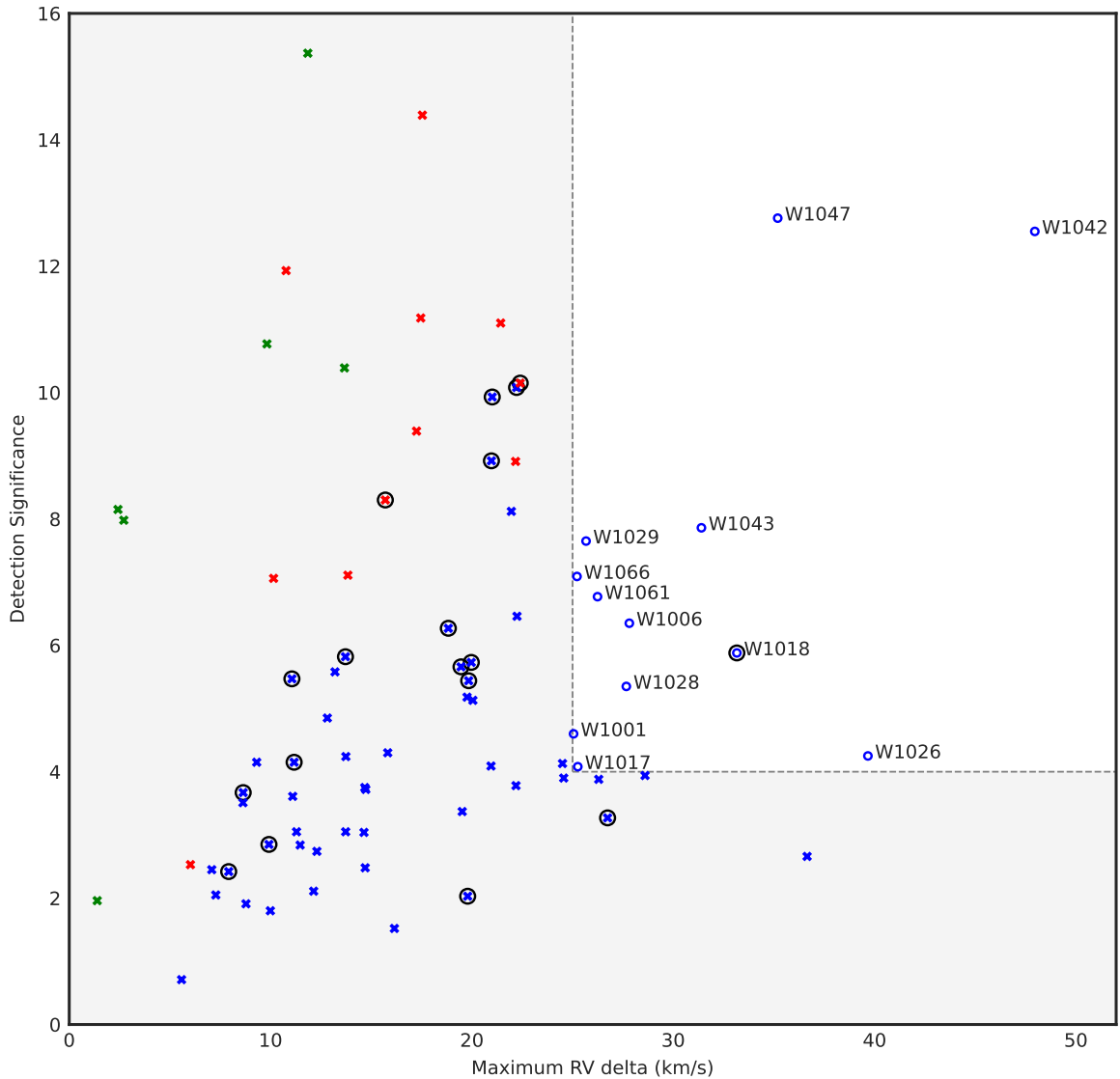}}
  \caption{$\Delta$RV vs. $\sigma_\textrm{det}$ for \textit{candidate} binaries. Markers follow the format of Figure~\ref{fig:rv-sigma}.}
\label{fig:rv-candidates}
\end{center}
\end{figure*}

\section{Candidate binaries}
\label{sec:probable}
Alongside the targets discussed in Section~\ref{sec:individual}, we find 12 systems with
$25<\Delta\textrm{RV}<50$\kms and $4<\sigma_{\textrm{det}}<15$. We label these objects as
\textit{candidates} based on the difficulty in establishing unambiguous orbital parameters with the
available data: we note that significant velocity changes nevertheless make it likely that these are
indeed \textit{bona fide} binary systems, especially for the five objects with
$\Delta\textrm{RV}\ge$30\kms, which is hard to reconcile with a non-binary interpretation.
Unsurprisingly, all candidates are O-type stars of luminosity class Iab-III: all B supergiants that
meet the $\Delta\textrm{RV}>25$\kms requirement also have $\sigma_{\textrm{det}}>15$, due to their
higher luminosity and narrow Paschen series lines. Only one of these candidate systems has more than
ten observations, and it is therefore likely that we have not sampled the full range of RVs for
these objects, especially if they have significant eccentricity: \object{W10} and \object{W1030} are
salient examples, appearing to be an unremarkable, low semi-amplitude systems at most epochs. 

\subsection{O9--9.5 I--III systems}
Four O9.5~I--III stars, one O9-O9.5~III star, and one O9~III star are identified as candidate
binaries. \object{W1018} (O9.5~Iab) is a hard X-ray source that displays a very similar $I$-band
spectrum to \object{W1027}. RVs spanning more than 30\kms appear incompatible with a non-binary
interpretation, but only four epochs of data are available, precluding more detailed examination.
\object{W1042}, \object{W1043} and \object{W1047} display strong similarities to the O9.5~II targets
discussed in Section~\ref{sec:brightgiants}. \object{W1042} (O9.5 II) is very likely to be binary,
with a RV range of $\sim$48\kms, while \object{W1043} (O9.5~II--III) also displays RV amplitudes
above 30\kms, but limited observations (eight and six epochs respectively) preclude accurate
determination of orbital parameters for either system. A FORS2/MXU $R$-band spectrum of
\object{W1043} is available, with H$\alpha$ appearing somewhat broad and infilled compared to
\object{W1022} and \object{W1050}, although He~I absorption lines appear single. \object{W1047}
(O9.5~II) was sampled on six epochs in 2008--2009 and a further four in 2013, and observations are
consistent with a $\sim5$~day period and near-circular orbit. A FORS2/MXU spectrum shows H$\alpha$
strongly in absorption while He~I again appears single-lined. All four systems represent strong
candidates for follow-up observation, which would allow robust orbital solutions to be determined. 

Finally, \object{W1026} (O9-9.5~III) and \object{W1066} (O9~III) show broader lines that imply a
lower luminosity class, with consequently increased fitting errors. \object{W1026} displays RV
changes of almost 40\kms, with significant changes in line width suggesting a secondary of similar
luminosity may be present. \object{W1066} shows lower RV variability, but $R$-band observations of
\object{W1066} suggest He~I~\lam7065 may be double. Both systems were only observed on six epochs in
2008--2009 and no further analysis can be carried out.   

\subsection{Broad-lined O-type stars}
\label{sec:broad}
Six targets (\object{W1001}, \object{W1006}, \object{W1017}, \object{W1028}, \object{W1029}, and
\object{W1061}) display broad Pa-11 and Pa-12 lines, and an otherwise featureless $I$-band spectrum.
All are low-luminosity targets (with $I\gsim15$), and all are classified as O+O? or O9-9.5~III~bin?
in Paper~III. None of these targets are detected at X-ray wavelengths (Paper~II), although a
FORS2/MXU spectrum of \object{W1028} is available, with H$\alpha$ appearing infilled and
He~I\lamlam6678,7065 possibly double.
 
 All five objects display RV changes that marginally exceed our criteria for classification as
 \textit{candidate} binaries, although none show RV amplitudes larger than $\sim$30\kms and all lie
 in the range $4<\sigma_{\textrm{det}}<8$. If these objects are indeed binary, then the RV changes
 reflect line-profile changes in heavily-blended Paschen series linßes; some support for this
 hypothesis is given by epoch-to-epoch variations in line width in \object{W1006}, \object{W1017},
 and \object{W1061}, although low $S$/$N$ limits the analysis of line-profile variability. Follow-up
 observations at other wavelengths will be required to confirm the nature of these objects, but we
 note the strong morphological similarity between these systems and other confirmed binaries (e.g.
 \object{W1002} or \object{W1021}) suggests that a binary interpretation is appropriate. 

 \section{Wolf Rayets}
 \label{sec:wr}
 The 2011 FLAMES campaign was focused on sampling the WC and WN population of Wd1, motivated by the
 analysis and determination of orbital parameters for \object{W13} (WNVL) and
 \object{W239}/\object{WR~F} (WC9d) described in \cite{ritchie10} and \cite{clark11}, and the
 observations of \object{W72}/\object{WR~A} (WN7b) and \object{WR~B} (WN7o+O) presented by
 \cite{bonanos07} and \cite{kb12}. This population traces the final stages of binary-mediated
 evolution in Wd1 from the post-interaction state (e.g. \object{W13}) to post-supernova disruption
 (\object{W5}/\object{WR~S}). 
 
 \subsection{WN Stars}
 \begin{figure}
  \begin{center}
    \resizebox{\hsize}{!}{\includegraphics{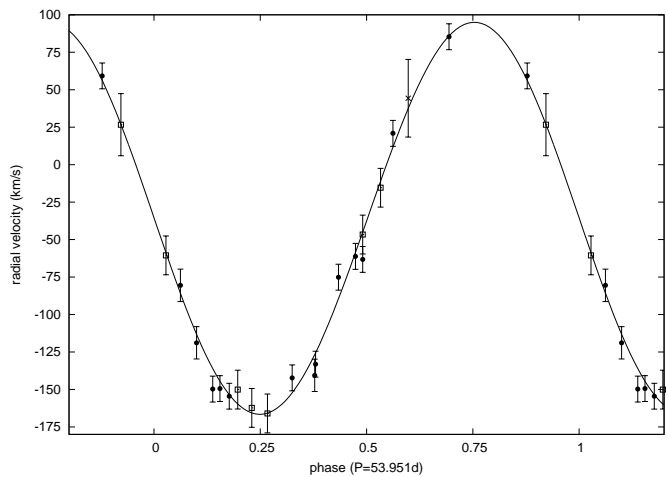}}
    \caption{RV curve for \object{W44}/\object{WR~L}. Open squares represent VLT/FLAMES observations from 2005, 2011, and 2013, filled circles represent VLT/UVES observations from 2016, and the point labeled with an $\times$ was derived from a low-resolution VLT/FORS spectrum obtained in 2004.}
    \label{fig:w44rv}
  \end{center}
 \end{figure}

\object{W44}/\object{WR~L} (WN9h; \citealt{crowther06}) was observed in HR21 mode on two nights in
2011 and four nights in 2013, with 14 epochs of data subsequently obtained with VLT/UVES in 2016. An
archival FLAMES spectrum in LR8 mode from 2005 and a VLT/FORS spectrum from 2004 were also included
in the analysis. RVs were measured using He~I\lamlam8845,8777,8583 absorption lines \citep{vh72},
and we find a period of $53.952\pm0.004$~days, with $K_1=130.8\pm3.7$\kms and
$\gamma=-36.4\pm2.3$\kms. The orbit is circular. \object{W44}/\object{WR~L} displays dramatic
emission line profile variability in H$\alpha$ and He~I \citep{clark10}, which are a blend of static
nebular emission and moving features that follow the WN9h star. The companion is not directly
observed, although if we assume a $20$\Msun mass for the emission-line
object\footnote{\cite{crowther06} suggest $18$\Msun for the WN7b star \object{WR~A}, while the WNVL
emission-line object in \object{W13} has a dynamical mass of 23\Msun \citep{ritchie10}.} and
$i\ge60^\circ$ based on the large semi-amplitude then its mass must be $\sim32-42$\Msun. A full
analysis will be presented in a forthcoming paper.  

 Although observations of hydrogen-rich, late-type WN stars in Wd1 have been very successful (cf.
 \citealt{ritchie10,clark14}), the remaining hydrogen-free WN stars cannot be analysed in
 similar detail. In HR16 mode, we only see weak lines of He~I\lam7065 in later types and
 N~IV\lam7114 in earlier types, but both features are broad and flat-topped, making them unsuited to
 RV measurement. In HR21 mode, only a weak He~II line is observed, blended with a telluric band. In
 two cases (\object{WR~J} and \object{WR~U}) we observe Paschen and N~I features that result from
 blending with nearby A-type hypergiants (\object{W12a} and \object{W16a} respectively), but we do
 not see any indicators of OB companions in $I$-band spectra of the remaining WN population, despite
 both \object{WR~A} and \object{WR~U} displaying strongly enhanced X-ray emission that would imply
 the presence of an O-type secondary. 
 
 \subsection{WC Stars}
 \object{WR~H} (WC9d) displays strong Paschen series absorption lines, most likely due to blending
 with the nearby luminous B1~Ia star \object{W46a}. \object{WR~N} (WC9d) also displays weak Paschen
 series features superimposed on dilute C~III and C~IV emission lines, but is located far to the
 south of the cluster where blending is less likely. It is tempting to associate the Paschen series
 features with a massive secondary, given the expectation that an OB companion is required for dust
 formation \citep{crowther06}, but confirmation would require additional observations. 

 Of the other WC targets, \object{W66}/\object{WR~M} (WC9d) was observed on four nights in
 2011 and displays RVs spanning approximately 100\kms, confirming it as binary, although
 the limited observations preclude further analysis. Variations in line profile are apparent at some
 epochs. \object{W241}/\object{WR~E} (WC9, observed on four nights) and \object{WR~T} (WC9d,
 observed on three nights) also display statistically-significant changes in RV, although with
 amplitudes below 25\kms these cannot be distinguished from wind variability. \object{WR~C} (WC9d)
 does not display significant variability. However, we caution that observations of these objects
 are very limited, and all four targets are appropriate for follow-up to extend the analysis of
 \cite{clark11}.

 \object{W39b}/\object{WR~K} (WC8) was observed in 2008--2009 and displays no sign of orbital
 motion, although only very large orbital changes would be apparent in its very broad and
 flat-topped C~III and C~IV emission lines. 

\section{Late-type targets}
\label{sec:midb}
\subsection{Luminous B1.5--B4 supergiants}
In addition to the already-discussed \object{W2a}, \object{W23a}, and \object{W71}, three other very
luminous early-mid B supergiants were observed in our study, including \object{W8b} (B1.5~Ia),
\object{W70} (B3~Ia), and \object{W57a} (B4~Ia). Their high luminosity and strong Paschen series
lines result in very precise RV measurement, and all three objects display substantial changes on a
timescale of days, although none exceed $\Delta\textrm{RV}>25$\kms. No clear periodicity or
secondary indicators of binarity are present, noting the difficulty in separating low semi-amplitude
motion from pulsational variability. As discussed in Paper~I, these objects appear to lie in an
instability strip that begins at $\sim$B0~Ia and grows in amplitude and period towards later
spectral types: we speculate that this instability may also be associated with the end of the
continuous spectral sequence from O9~III to B4~Ia that we observe in Wd1, delineating a more rapid
transition to the A--F hypergiant population. Given the importance of these high-amplitude pulsators
in setting the RV threshold for robust identification of binary systems, further observations are
required to fully characterise their pulsational behaviour. 

\subsection{The B hypergiants W1049 and W1069}
\object{W1049} and \object{W1069} (B1-2~Ia$^+$ and B5~Ia$^+$ respectively; Paper~III) are
newly-identified B hypergiants associated with Wd1. The two objects were each observed on 12 epochs
in 2013, with multiple observations being obtained on two nights: in these cases, measured RVs
agreed to within 1\kms. Both objects display line profile variability and low-level RV changes in
the Paschen series, which are likely due to pulsational instability, but larger RV changes
indicative of a binary nature are not observed.  

Both objects lie several arcminutes from the cluster core, with \object{W1049} $3'28''$ North of the
cluster centre, and \object{W1069} $4'15''$ to the South~East. Unlike the B1.5~Ia$^+$/WNVL cluster
member \object{W5} (=\object{WR~S}; \citealt{clark14}), neither object has a systemic velocity
strongly inconsistent with the cluster mean, although \object{W1049} shows an offset of
$\sim$15\kms. Given the suggestion that \object{W5} is the remnant of a short-period binary system
disrupted in a Type Ibc supernova \citep{clark14} and the expected supernova rate in Wd1
(Paper~III), it is tempting to ascribe a similar evolutionary pathway to \object{W1049} and
\object{W1069}, with Roche-lobe overflow and binary-mediated mass transfer leading to their current
overluminous WNLh/B~Ia$^+$ state and their ejection from the cluster in earlier supernova events. 

\subsection{A--M hypergiants}
The LBV \object{W243} was observed in both 2011 and 2013, extending the 2002--2009 baseline of
observations discussed by \cite{ritchie09b}. Classified as B2~Ia by \cite{w87}, A2~Ia by
\cite{clark04}, and A3~Ia$^+$ by \cite{ritchie09b}, the 2013 spectra suggest a A4--5~Ia$^+$
classification based on the strength of the temperature-sensitive N~I~\lam8680\dots8686 triplet,
which now appear directly comparable to \object{W16a} (A5~Ia$^+$; \citealt{cncg05}). In addition,
$I$-band Ca~II and Mg~I emission lines have weakened to the extent that they can be barely
distinguished in 2013 spectra. The limited coverage of FLAMES in HR21 mode precludes examination of
other temperature-sensitive features in the complex emission/absorption spectrum of \object{W243}.  

Five YHGs were observed in 2011 or 2013. Comparison with archival FORS and FLAMES spectra from
2004--2005 precludes \textit{significant} evolution in \object{W4} (F3~Ia$^+$), \object{W12a}
(F1~Ia$^+$), or \object{W16a} (A5~Ia$^+$) during the intervening period, while \object{W8a}
(F8~Ia$^+$; \citealt{cncg05}) appears as $\sim$F5~Ia$^+$ in 2013 spectra. \object{W265}
(F1--5~Ia$^+$) was discussed in detail in \cite{clark10}, with further observations in 2011 and 2013
supporting the presence of a slow pulsational cycle with quasi-period $\sim$10$^2$~days accompanied
by changes in temperature-sensitive lines. However, our current baseline shows no evidence for
long-term secular evolution in any of these YHGs, such as that seen in \object{$\rho$~Cas},
\object{HR8752}, or \object{IRC+10~420}.  

Finally, the RSGs \object{W75} and \object{W237}  were observed in the 2013 campaign. \object{W75}
appears as $\sim$M0~Ia based on the weakness of the TiO\lam8860 bandhead, consistent with
earlier observations described by \cite{clark10}. \object{W237} displays a slightly stronger TiO
bandhead in 2013 than in archival FLAMES observations from 2005, and we suggest a M3--4~Ia
classification, noting the lower resolution of the earlier spectra. Variability is negligible over
the $\sim$4-week baseline of observations, with epoch-to-epoch observations in agreement to better
than 1\kms. We note that the RV of $-47.9\pm0.3$~km/s obtained for \object{W237} is in close
agreement with the value obtained from H$_2$O maser profiles by \cite{fok}.

\section{Notable non-detections}
\label{sec:non}
Finally, we turn to targets that display secondary indicators of binarity, but are not identified as
RV binaries or candidate binaries. These are predominantly X-ray sources (\citealt{clark08} and
Paper~II) and stars with spectral morphologies suggesting a binary nature (Paper~III), but also
include non-thermal radio sources (\citealt{dougherty}, \citealt{fenech}, and \citealt{andrews}). 

\subsection{X-ray sources}
Of all the luminous X-ray sources discussed in Paper~II, only \object{W27}, an O7--8~Ia$^+$ blue
straggler discussed by \cite{clark19a}, and the WN6o star \object{WR~U} \citep{crowther06} are not
identified as binaries or candidate binaries. As discussed in Section~\ref{sec:wr}, none of our
FLAMES configurations provide features suitable for RV measurement in a hydrogen-free WN star,
although the strong similarities with the X-ray luminous WN7o~+~O stars \object{WR~A} and
\object{WR~B} \citep{bonanos07,clark08} nevertheless suggest \object{WR~U} is also a WN~+~O binary.
\object{W27} was only observed on one night in 2011 and three nights in 2013. No large excursions in
RV are observed, although we note that the mean RV is offset from the cluster mean by $\sim$20\kms,
potentially indicating incomplete sampling of more significant RV changes. Secondary indicators of a
binary nature include strong X-ray emission (Paper~II), a complex H$\alpha$ profile, and a
potentially-double He~I~\lam6678 line \citep{neg10}, and this object also remains a strong binary
candidate. 

Although enhanced X-ray emission is a strong indicator of binarity \citep{skinner, pittard}, it is
notable that few of the sources with $L_X\sim10^{32}$~erg~s$^{-1}$ discussed in Paper~II have been
confirmed as spectroscopic binaries in this work: 17 objects do not display
statistically-significant RV variability or have $\Delta\textrm{RV}<25$\kms, while only
\object{W6a}, \object{W10}, \object{W24}, \object{W232}, and \object{W1027} are confirmed binaries,
and \object{W1018} is a candidate binary. \cite{clark08} identified eight OB supergiants with
$kT\ge1.4$~keV as likely colliding-wind binaries, with two (\object{W13} and \object{W24})
subsequently confirmed as short-period systems containing two post-MS objects (\citealt{ritchie10}
and this work). Of the remainder, \object{W47} (O9.5~Iab) has very similar properties to
\object{W24}, and was observed on three nights in 2013. \object{W65} (O9~Ib) and \object{W1055}
(B0~Ib) are harder and less luminous sources. Both were observed on six nights in 2008--2009, while
\object{W65} was observed on a further three nights in 2013. Although all three objects display
statistically-significant RV changes, none meet our $\Delta\textrm{RV}>25$\kms criterion, although
we caution that \object{W47} in particular has too few observations to draw definitive conclusions;
like \object{W27}, its mean RV is significantly offset from the cluster mean, again suggesting
incomplete sampling. In addition, the $I$-band spectra of \object{W65} and \object{W1055} are
suggestive of possible binarity. In the case of \object{W65}, the Paschen lines are typical of
lower-luminosity O9 stars, with Pa-11 very similar to \object{W1015} (O9~III), but the C~III/Pa-16
blend is broad and weak, and the star seems rather too luminous for its early spectral type.
Similarly, \object{W1065} displays stronger C~III than expected for its B0~Ib spectral type,
suggesting the presence of an O-type companion. \object{W56b} (O9.5~Ib), \object{W1036} (=C07-X4;
O9.5~Ib), and \object{W1041} (O9.5~Iab) all have $kT\sim1.6$keV but do not display
statistically-significant RV changes, although \object{W1041} appears to be a potential binary from
its $I$-band morphology and was noted as a candidate binary in Paper~III. Of the X-ray sources with
$kT<1$keV, \object{W1033} (=C07-X5; O9--9.5~I--III) and \object{W1040} (=C07-X3; O9--9.5~I--III)
both seem somewhat too luminous for their apparent spectral type, but other objects such as
\object{W1005} (=C07-X7; B0~Iab), \object{W1051} (O9~III), and \object{W1064} (O9.5~Iab) show no
indications of binarity other than the moderately-enhanced X-ray emission noted by \cite{clark08}.

The lack of strong RV variability in these objects may represent unfavourable inclination rather
than isolated stars, and we note that the 6.60~day hard X-ray source \object{W24} and the X-ray
luminous O4--6~Ia$^+$ interacting binary \object{W30a} also have very low semi-amplitudes, despite
their massive components and compact configuration. Others may be substantially eccentric systems
(cf. \object{W10}), with enhanced X-ray emission only near periastron; our survey has limited
sensitivity to such a configuration, especially when only a few epochs of data are available.
Alternatively, some of these objects may simply be isolated OB stars with X-ray luminosities
consistent with the empirical $L_\textrm{x}\sim10^{-7}L_\textrm{bol}$ relationship, implying that
X-ray detection alone is insufficient for binary classification unless emission is substantially
enhanced (cf. the five OB and three WR stars with $L_x\gsim10^{33}$~erg~s$^{-1}$ identified by
\citealt{clark08}) or other binary indicators (e.g. composite spectra, or hard X-ray emission
indicative of formation in a wind-collision zone) are present.

\subsection{Morphology}
With the exception of \object{W27}, which is too poorly-sampled to meet our criteria for binary
classification, all of the luminous OB objects identified as candidate binaries by \cite{neg10}
based on $R$- and $I$-band spectra are confirmed as binaries in this work. In contrast, Paper~III
identifies 25 lower-luminosity objects as candidate O-type binaries based on their $I$-band
morphology (classified as either O9-9.5~III~bin? or simply as O+O?), but we are only able to confirm
two of these objects (\object{W1002} and \object{W1013}) as binary in Section~\ref{sec:individual},
while a third (\object{W1021}) was identified as an eclipsing binary by \cite{bonanos07}. A further
six are identified as candidate binaries in Section~\ref{sec:probable}, in every case only just
exceeding the $\Delta\textrm{RV}>25$\kms threshold. None of these objects are listed in Paper~II,
suggesting they are indeed lower-luminosity objects with weak winds that cannot drive detectable
X-ray emission. 

The O9~III stars \object{W6b} and \object{W1063} demonstrate that recovery of orbital properties is
possible for the faintest targets in our survey, but these objects appear to have lower-luminosity
OB~IV-V secondaries, implying an unequal mass ratio, and appear as unblended, single-lined stars. In
contrast, the morphological similarities of the broad-lined systems suggests that we may be
observing a substantial population of $\sim$O9~III+O9~III binaries, but only those at the most
favourable inclination display sufficient RV modulation to yield measurable line profile variations
in their blended Paschen-series lines. Consideration of synthetic spectra suggests that orbital
modulation in line profile would be sufficient to meet the criteria in Section~\ref{sec:cands} for
$i\gsim60^\circ$ and $P\lsim10$~days, but the broad Paschen series lines in a short-period
O9~III+O9~III system would remain heavily blended, precluding recovery of orbital parameters. Our
ability to accurately measure RV changes in these objects using the Paschen series lines is
therefore very limited. The high reddening towards Wd1 means that alternative photospheric lines are
largely unavailable, although studies of He~I\lam7065 represent a potential route forward with these
objects (cf. \citealt{kb12}). 

\subsection{Radio sources}
\cite{dougherty} reported non-thermal emission from two targets, \object{W15} (O9~Ib) and
\object{W17} (O9~Iab), although \cite{fenech} report detecting only \object{W17}. In addition, radio
emission from \object{W1031} (O9~III) and \object{W1056} (O9.5~II) is reported by \cite{andrews}.
\object{W15}, \object{W17}, and \object{W1031} do not meet the criteria for classification as
candidate binaries, although observations are limited to only six or seven epochs, and in all three
cases statistically-significant RV changes are present with amplitudes $\sim$20\kms, close to our
threshold for classification. Our observations are therefore consistent with the suggestion that
these systems are all long-period and potentially eccentric binaries \citep{andrews} that may be
hard to detect conclusively in an RV survey unless viewed at favourable inclination. \object{W1056}
is the only OB star that is both a radio detection and a confirmed binary, and although its orbital
parameters are not yet secure, our provisional period of 53.1~days with $e\sim0.3$ would be in good
agreement with such a scenario. The status of \object{W1030}, another long period, eccentric system
that likely contains an O9~Ib (or possibly B0~Ia) primary and post-MS secondary is uncertain: it
lies outside the coverage reported by \cite{andrews}, and is not detected by \cite{fenech}.  

\section{Discussion}\label{sec:discussion}
\begin{table*}[!htbp]
  \caption{Binaries in Westerlund~1, with orbital solutions where available. Errors are omitted to save space, but are given in the text or in the cited references. Values where the orbital period is candidate or tentatively identified are marked with one (candidate) or two (tentative) question marks, while other parameters with errors $\ge20\%$ are also highlighted with a question mark.}
  \label{tab:results}
  \begin{center}
  \begin{tabular}{lllcccccc}
                &                   &             & $P$  & $K_1$   & $K_2$   & $\gamma$ & $e$        & $\omega$ \\
  ID            & Spectral Type     & Type$^\dag$ &(days) & (\kms)  &  (\kms) & (\kms)   &            & (degrees)\\
  \hline\hline
  &&&&&\\
  W2a           & B2~Ia             & SB1      & $44.95$??& $18.5$  & -       & $-44.6$  & $\sim0.3$? & $285$?\\
  W6a           & B0.5~Iab          & SB1/P, X & $2.21$?  & $25$?   & -       & $-38.9$  & $0$        & - \\
  W6b           & O9.5~III          & SB1      & $2.65$?? & $41.6$  & -       & $-64.2$  & $0.29$?    & $249$? \\
  W10           & B0.5~Ia~+~O9.5~II & SB2, X   & -        & $75$?   & $85$?   & $-34.7$  & $\ge0.3$?  & $0$?\\
  W23a          & B2~Ia + O9~I      & SB1$^\star$    & $40.46$? & $9.2$?   & -       & $-52.5$  & $\sim0.2$? & $220$?\\
  W24           & O9~Iab + O9~I?    & SB1$^\star$, X & $6.60$?  & $13.6$  & -       & $-50.4$  & $0.25$?   & $212$\\
  W30a          & O4--6~Ia$^{+}$ + O& SB1, X   & $6.3$?   & $13$?   & -       & $-42.3$  & $0$        & - \\
  W36$^{a,b}$   & OB~+~OB           & SB2/E, X & $3.18$   & $175$   & $235$   & $-37$    & $0$        & - \\
  W43a          & B0~Ia             & SB1      & $16.23$  & $69.4$  & -       & $-64.4$  & $0.10$     & $149$ \\
  W52           & B1.5~Ia           & SB1/P    & $6.62$?  & $24.7$? & -       & $-48$?   & $0$?       & - \\
  W53a          & O~+~O             & SB2/P, X & $1.3$    & $125?$  & $140?$  & $-40?$   & $0$        & - \\
  W71           & B2.5~Ia           & SB1      & $36.64$??& $9.8$   & -       & $-44.8$  & $\sim0.1$? & $90$?\\ 
  W232          & B0~Iab            & SB1      & $9.98$   & $14.7$? & -       & $-45.3$  & $0.41$     & $256$?\\
  W1021$^{a,b}$ & O~+~O             & SB2/E    & $4.45$   & $229$   & $187$   & $-40$    & $0.18$     & $252$ \\
  W1022         & O9.5~II           & SB1      & $6.18$?  & $73.2$? & -       & $-36.6$? & $\le0.3$?      & -\\
  W1030         & O9.5~Ib           & SB1      & $173.85$ & $16.5$  & -       & $-46.6$  & $0.56$     & $323$\\
  W1048         & O9.5~Ib           & SB1/E?   & $8.75$   & $76.2$  & -       & $-64.0$  & $0.08$     & $205$\\
  W1050         & O9.5~II           & SB1      & $4.06$?  & $27.6$? & -       & $-51.0$  & $0.56$?    & $161$\\
  W1056         & O9.5~II           & SB1      & $53.13$? & $28.7$  & -       & $-41.8$  & $0.31$     & $55$?\\
  W1060         & O9.5~II           & SB1      & $3.24$   & $28.7$? & -       & $-59.3$  & $\le0.1$?  & -\\
  W1063         & O9~III~+~O?       & SB1      & $15.44$? & $38.6$  & -       & $-47.8$  & $0.32$?    & $233$\\
  W1065         & B0~Ib             & SB1      & $11.13$  & $52.8$  & -       & $-39.4$  & $0.38$     & $256$\\
  W1067         & B0~Ib             & SB1      & $6.13$   & $13.2$? & -       & $-46.9$  & $0.52$?    & $0$?\\
  \\ 
  W13$^{a,b,c}$     & WNVL~+~OB     & SB2/E, X & $9.27$   & $210.2$ & $137.3$ & $-48.2$  & $0$        & - \\
  WR~A / W72$^a$    & WN7b~+~O?     & P, X     & $7.63$   & -       & -       & -        & -          & - \\
  WR~B$^{a,b}$      & WN7o~+~O?     & SB1/E, X & $3.52$   & -       & $360$   & $-35$    & $0$        & - \\
  WR~F / W239$^{d}$ & WC9d~+~O~+~?  & SB1      & $5.05$   & $39.7$  & -       & $-60.5$  & $0$        & - \\
  WR~L / W44        & WN9h:~+~O?    & SB1      & $53.95$  & $130.8$ & -       & $-36.4$  & $0$        & - \\
  \end{tabular}
  \end{center}
  Notes: $^\dag$Listed as Single/Double lined ({\bf SB1}/{\bf SB2}) and {\bf E}clipsing
  or {\bf P}eriodic photometric variables. Strong {\bf X}-ray sources are also noted (\citealt{clark19a}).
  $^\star$The influence of C~III$\lambda$8500 absorption originating in a luminous O9 secondary is apparent, but lines from the secondary cannot be observed directly, and the object is therefore listed as an SB1.\\
  References: $^a$\cite{bonanos07}, $^b$\cite{kb12}, $^c$\cite{ritchie10}, $^{d}$\cite{clark11}. 
\end{table*}

\begin{table*}[!htpb]
  \caption{Candidate Binaries in Wd1, listing indicators of binarity: RV changes, X-Ray Luminosity, Infra-Red Excess, Non-thermal Radio Emission, Spectral Morphology (e.g. Double Lines, Broad Lines, or anomalously Weak Lines). Indicators that are strongly inconsistent with a single star are shown as solid circles, while indicators that suggest a binary interpretation are shown as open circles. Objects are listed if they (a) display substantial RV changes ($\Delta\textrm{RV}>25$\kms), (b) display any single secondary indicator strongly inconsistent with a single star, or (c) display two or more indicators that individually suggest binarity.}
  \label{tab:candidates}
  \begin{center}
  \begin{tabular}{llccccc}
  ID         & Spectral Type$^{a}$ & Radial Velocity & X-Ray$^b$ & Infra-Red & Radio & Morphology\\
  \hline\hline
  &&&&\\
  W9       & sgB[e]             &         & \y       & \y$^c$    &  $\circ^c$  &\\
  W15      & O9~Ib              &         & $\circ$  &           & $\circ^d$   &\\
  W17      & O9~Iab             & $\circ$ & $\circ$  &           & \y$^{d}$    &\\
  W27      & O7--8~Ia$^{+}$     & $\circ$ &  \y      &           &             &\\
  W47      & O9.5~Iab           & $\circ$ & $\circ$  &           &             &\\
  W65      & O9~Ib              & $\circ$ & $\circ$  &           &             & $\circ$\\
  W1001    & O~+~O?             & \y      &          &           &             & $\circ$\\
  W1002    & O9~II--III~+~O?    & \y      &          &           &             & \y \\
  W1006    & O9--9.5~III~bin?   & \y      &          &           &             & $\circ$\\
  W1013    & O~+~O?             & \y      &          &           &             & $\circ$\\ 
  W1017    & O9--9.5~III~bin?   & \y      &          &           &             & $\circ$\\
  W1018    & O9.5~Iab           & \y      & $\circ$  &\\
  W1026    & O9--9.5~III        & \y      & \\
  W1027    & O9.5~Iab           & \y      & $\circ$  &           &             & $\circ$\\
  W1028    & O9--9.5~III~bin?   & \y      &          &           &             & $\circ$\\
  W1029    & O9--9.5~III~bin?   & \y      &          &           &             & $\circ$\\
  W1031    & O9~III             &         & $\circ$  &           & $\circ^e$   &\\
  W1033    & O9-9.5 I--III      &         & $\circ$  &           &             & $\circ$\\
  W1040    & B0~Iab             &         & $\circ$  &           &             & $\circ$\\
  W1041    & O9.5~Iab + ?       &         & $\circ$  &           &             & $\circ$\\
  W1042    & O9.5~II            & \y      & \\
  W1043    & O9.5~II--III       & \y      & \\
  W1046    & O~+~O?             &         &          &           &             & \y$^g$\\
  W1047    & O9.5~II            & \y      & \\
  W1055    & B0~Ib~(+O?)        &         & $\circ$  &           &             & $\circ$\\
  W1061    & O9--9.5~III~bin?   & \y      &          &           &             & $\circ$\\
  W1066    & O9~III             & \y      &          &           &             & \y\\  
  WR~C     & WC9d               &         &          & \y$^f$ \\
  WR~H     & WC9d               &         &          & \y$^f$\\
  WR~M     & WC9d + O?          & \y      &          & \y$^f$\\
  WR~N     & WC9d               &         & $\circ$  & \y$^f$   &              & $\circ$\\ 
  WR~T     & WC9d               &         &          & \y$^f$\\
  WR~U     & WN6o               &         & \y       & \\
  \hline
  \end{tabular}
  \end{center}
  References: $^a$\cite{clark20}, $^b$\cite{clark19b}, $^c$\cite{clark13},$^d$\cite{dougherty}, $^e$\cite{andrews},  $^f$\cite{crowther06}, $^g$Paper~III.
\end{table*}

\subsection{Sensitivity}
This initial study of Wd1 with VLT/FLAMES has demonstrated the challenge of detecting binary systems
in the cluster. High reddening precludes the use of standard diagnostic lines in the blue region of
the spectrum, particularly the strong He~I, He~II, and Si~III lines covered by FLAMES setups
LR02--LR03 and HR02--HR06, and we are forced to rely on Paschen series lines in setup HR21. These
are most effective for single-lined $\sim$O9.5~II-B0~Ia systems, where luminosity (and hence
$S$/$N$) is increasing rapidly but pulsational variability is less pronounced. Although the Paschen
lines are very strong in later types, disentangling orbital and pulsational variability becomes
increasingly challenging, while early-type systems suffer from broad lines and low $S$/$N$. We have
very limited sensitivity to double-lined systems at any spectral type, as separation of the Paschen
series lines requires the system to have a large semi-amplitude, coupled in some cases with
fortuitous timing (e.g. \object{W10}). We would expect the supergiant population to skew towards
SB1s, as the high luminosity of the primary would imply the secondary is undetectable unless the
mass ratio is close to unity and both systems are in a similar evolutionary state. Double-lined
systems become much more likely at lower luminosity classes, and it is likely that we are observing
a large population of heavily blended SB2s in the O9~III population. We briefly summarise our
sensitivity as follows: 
\begin{itemize}
  \item At short orbital periods ($P<5$~days), the most massive stars will have already undergone
  Roche-lobe overflow, and even under favourable inclinations our ability to separate the Paschen
  series lines in short-period interacting systems (e.g. \object{W36} or \object{W53a}) is very
  limited. Pre-interaction systems may be found at $\sim$O9.5~II and earlier (e.g. \object{W1021} or
  \object{W1050}), and orbital parameters can be recovered if the system is an SB1.   
  \item Sensitivity peaks for SB1s with spectral types in the range O9.5~II to B0~Ia and periods
  $5<P\lsim15$~days. Such systems represent an optimum balance of relatively high luminosity, narrow
  Paschen series features, lower pulsational instability, and orbital periods that make it likely
  that we will obtain reasonable sampling of the full RV curve. The majority of systems
  identified as binary fall in this range. 
  \item At longer orbital periods ($P>20$~days), sensitivity drops rapidly and a large number of
  observations are required to separate orbital and photospheric variability; it is also
  increasingly likely that we do not have full sampling of the full range of RVs. Objects later than
  $\sim$B1~Ia require wider orbits to accommodate the primary, but growing pulsational instability
  requires favourable inclination and/or a luminous secondary (cf. \object{W23a}) to unambiguously
  support a binary interpretation.  
  \item At $P\gsim100$~days sensitivity is very low, with \object{W1030} representing an almost
  ideal case of favourable spectral type (O9.5~Ib) and a relatively large number of observations
  spanning a five-year baseline. We would be unable to conclusively detect this object at earlier or
  later spectral types.  
  \item We have limited sensitivity to systems with high eccentricity, which is reflected in the
  lack of any robust solutions with $e\gg0.5$. RVs may be below the $\Delta\textrm{RV}>25$\kms cut
  for much of the orbit in highly-eccentric systems, and even when a discrepant RV near periastron
  confirms the object as a binary, the orbital solution will be unreliable unless the periastron
  passage is well sampled. 
\end{itemize}
Many of the identified binary and candidate binary systems exceed $\Delta\textrm{RV}>25$\kms in the
2008--2009 dataset alone, implying that a $\Delta$RV cut can effectively detect (candidate) binaries
even if the number of epochs of observation is low. However, the additional RV data and extended
baseline of observations provided by the 2011 and 2013 observations greatly supported determination
of orbital parameters, and few of the objects discussed in Section~\ref{sec:individual} had robust
solutions based on just the 2008--2009 dataset. 

\subsection{Cumulative Distribution of orbital periods}
\begin{figure}
  \begin{center}
    \resizebox{\hsize}{!}{\includegraphics{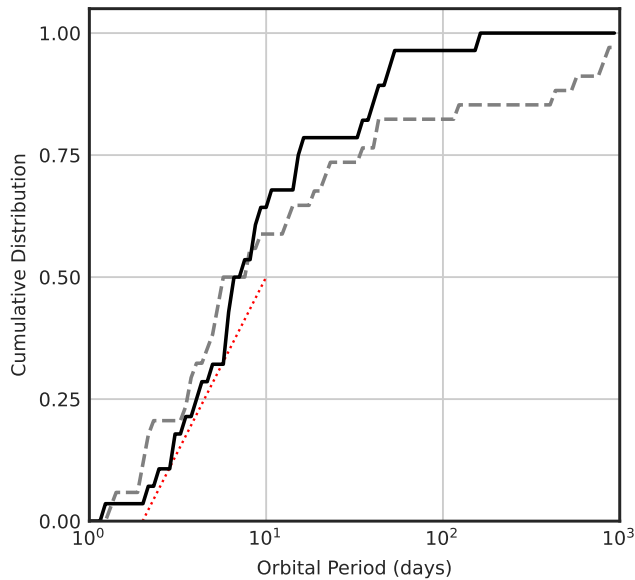}}
    \caption{Preliminary cumulative distribition of orbital periods from the OB stars in this work (solid line) and O stars in galactic clusters (\citealt{sana12}; dashed line). The dotted line shows the expected distribution if half the binary systems have a period in the range 2--10~days (\citealt{sana11}).}
    \label{fig:cumdist}
  \end{center}
\end{figure}
Figure~\ref{fig:cumdist} plots a preliminary cumulative distribution function (CDF) of orbital
periods for Wd1, with the sample of Galactic O~stars from \cite{sana12} plotted for comparison,
noting that we include candidate periods and make no attempt to correct for bias, and further work
is required before this result can be considered robust. Nevertheless, Wd1 appears to share the
overabundance of short-period systems noted in other massive clusters (e.g.
\citealt{sana11,kobulnicky,almeida}), while the very steep gradient of the CDF between 5--15 days
reflects the peak sensitivity in this range noted in the previous section. The CDF flattens above
$\sim50$~days as a result of our insensitivity to long period systems. We tentatively note an
apparent lack of very short-period systems compared to \cite{sana12}, and note that all of the
luminous objects with $P\lsim5$~days are interacting systems (e.g. \object{W6a}, \object{W36}, or
\object{W53a}) or post-interaction systems (e.g. \object{WR-B} or \object{W239}/\object{WR~F}). At a
cluster age of $\sim5$Myr \citep{neg10} any massive ($\gsim40$\Msun) binaries with very short
orbital periods will have already undergone Roche-lobe overflow (e.g.
\citealt{wellstein01,petrovic}). At lower masses, very short period systems will still be in a
pre-interaction state, but will be heavily blended unless observed at favourable inclination, and
only a handful of early-type systems have confirmed (\object{W1021}, \object{W1060}) or candidate
(\object{W6b}, \object{W1050}) periods below five days. 

\subsection{The OB Binary Fraction in Westerlund~1}
Of the 166 confirmed cluster members listed in Paper~III, 133 are OB stars, with spectral types
ranging from $\sim$O9~III to B9~Ia$^+$; of the remainder, 22 are Wolf-Rayets, while 11 are
cool-phase hypergiants. We report 26 OB binaries in Section~\ref{sec:individual}, while
Section~\ref{sec:probable} adds a further 12 objects; although we classify these as
\textit{candidates} due to the challenge of extracting orbital parameters, they exceed the
$\Delta$RV$>25$\kms cut and are very likely to be binary (see also \citealt{sana13,sd20};
Lohr~et~al.~in~prep.). We estimate incompleteness due to unfavourable inclination at $\sim10\%$ in
Section~\ref{sec:cands}. Collectively, this suggests a minimum binary fraction in the OB population
of $\sim40\%$, consistent with the initial assessment in Paper~I. This value is below the O~star
binary fraction of $69\pm9\%$ in Galactic open clusters \citep{sana12}, but slightly higher than the
spectroscopic binary fraction of $\sim35\%$ reported for \object{30~Dor} by \cite{sana13}, which
corresponds to objects displaying $\Delta$RV$>20$\kms. A true binary fraction of $51\%$ is inferred,
although \cite{almeida} suggests $\sim60\%$ may be more appropriate. 

Several factors could increase our estimated binary fraction in Wd1. Firstly, a substantial
population of broad-lined O9~III objects were proposed as binary in Paper~III, and evidence
presented here supports this conjecture: if confirmed, the binary fraction must be in excess of
$50\%$. Secondly, we have very limited sensitivity to long period, eccentric, or unequal-mass binary
systems, and have identified only one system with $P\gg50$~days and none with $e\gg0.5$. If we
assume that Wd1 follows other Galactic clusters (cf. \citealt{sana12}), then we would expect a
significant number of currently-undetected systems, even at favourable spectral type. Thirdly, some
objects have very limited sampling and do not meet the $\Delta$RV$>25$\kms cut based on the
available data, but display an offset mean~RV or epoch-to-epoch variability that suggests that
additional observations would confirm their binary nature. Other OB cluster members have no
multi-epoch observations at all. Finally, we have objects with very strong secondary indicators of
binarity, such as the sgB[e] star \object{W9}, the blue straggler \object{W27}, or the non-thermal
radio source \object{W17}, that have not been confirmed as RV binaries in this paper. 

Although there is overlap between these categories, they collectively suggest that binary fraction
might even approach the $\sim70\%$ inferred for the WR population \citep{crowther06}. However, an
equivalence would be surprising, as the Wolf-Rayet population is believed to be shaped by
binary-mediated evolution, and we caution that the target selection discussed in Paper~I may be
biased towards binary systems, while the 2011 and 2013 datasets preferentially select candidate
systems for follow-up observations. A conclusive determination of the cluster binary fraction must
therefore await a future programme of work that provides a substantial baseline of observation
across an unbiased sample of cluster objects.  

\subsection{Binary Interaction}
\begin{figure*}[htp]
  \begin{center}
    \resizebox{\hsize}{!}{\includegraphics{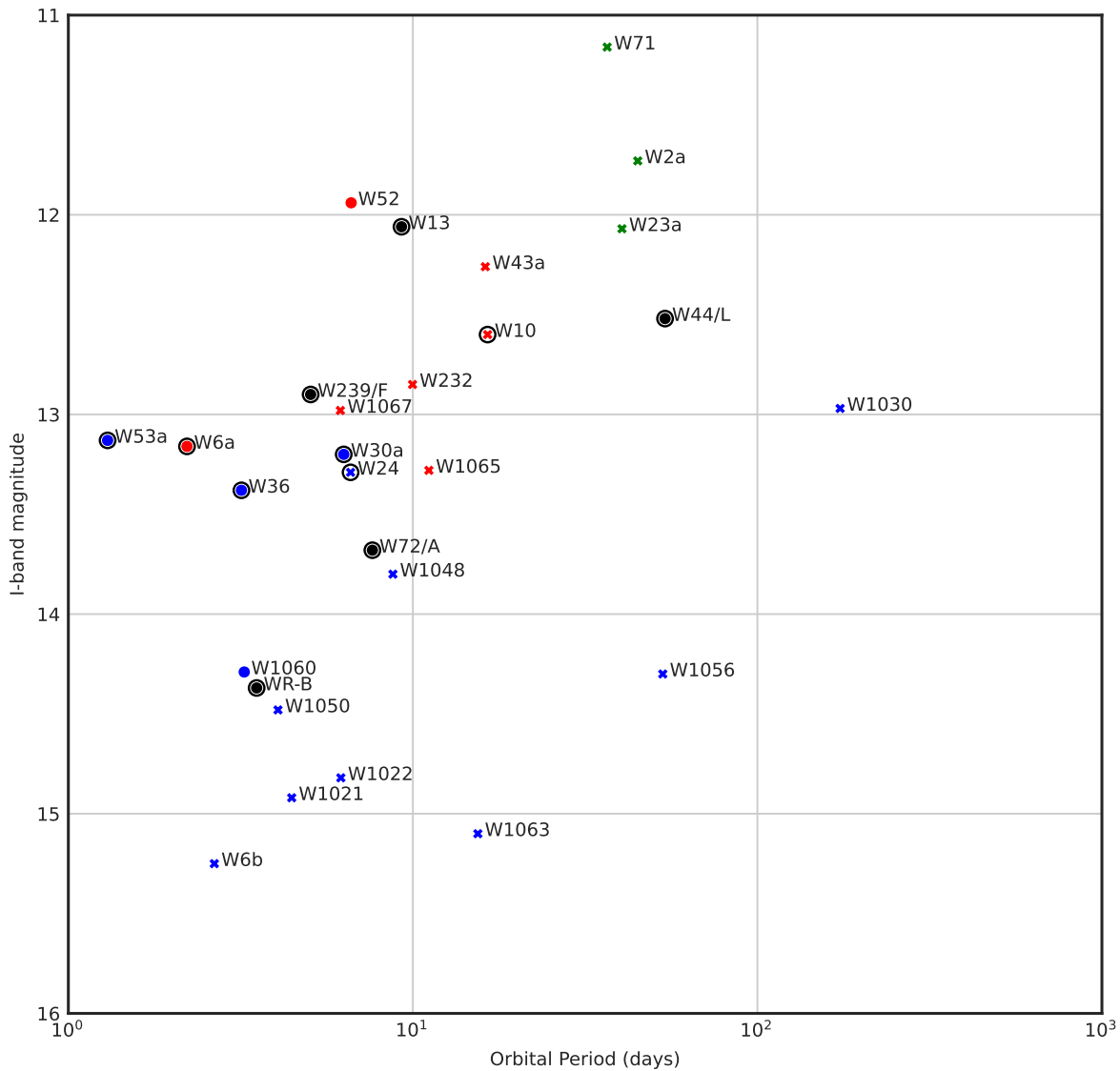}}
    \caption{Orbital period vs. $I$-band magnitude for target with photometry from \cite{cncg05} or \cite{bonanos07}. Systems with circular orbits are plotted as filled circles, while systems with eccentric orbits are plotted with `$\times$' symbols. Black circles show X-ray sources from Paper~II. Red, green, and blue points indicate spectral type, following Figures~\ref{fig:rv-sigma} and~\ref{fig:rv-candidates}.
    Note that the orbital period for \object{W10} is not firm, and its $I$-band magnitude is estimated from its $R$-band magnitude of 18.516 \citep{bonanos07} under the assumption that it has the same reddening as other nearby early-B supergiants.}
    \label{fig:Iper}
  \end{center}
\end{figure*}
Figure~\ref{fig:Iper} plots the orbital period of confirmed and candidate binary systems against
their $I$-band magnitude, which serves as a crude proxy for evolutionary state. Systems with zero
and non-zero eccentricity are plotted separately. Although a circular orbit does not necessarily
imply a post-interaction state, almost all targets with $P<10$~days and $e=0$ are either interacting
systems (\object{W6}, \object{W30a}, \object{W36}, or \object{W53a}) or post-interaction Wolf-Rayet
binaries (\object{W13}, \object{W72}/{WR~A}, \object{WR~B}, or \object{W239}/\object{WR~F}). The
evolutionary state of \object{W52} is unclear: given its high luminosity, a short period and
circular orbit suggest a post-interaction system, but $R$-band spectroscopy shows an unremarkable
B1.5~Ia supergiant with weak H$\alpha$ emission, while He~I is in absorption; the post-interaction
early-B/WNVL objects \object{W5}/\object{WR~S} and \object{W13} display very different spectra with
strong H$\alpha$ and He~I emission. Further observations are required to determine its state.
\object{W1060} is a less-evolved O9.5~II star with a very short period. 

Given that most luminous ($I<14$), short-period ($P<10$~days) systems have circular orbits,
\object{W24} (O9~Iab+O9~I?) and \object{W1067} (B0~Ib) stand out for retaining significant
eccentricity; lack of circularisation in the short-period O9~II-III systems with $I$-band magnitudes
fainter than 14 is less surprising, as these less-evolved systems are not yet expected to interact,
while \object{W1048} (O9.5~Ib) has $e<0.1$. \object{W1067} displays H$\alpha$ emission with a weak
P~Cygni profile, which is unusual for its spectral type and suggests a potentially-interacting
system, while the 6.6~day candidate period and near-equal mass ratio of \object{W24} also implies
that the primary must be close to Roche-lobe overflow. \object{W24} may therefore be a direct
precursor to the sgB[e] \object{W9} \citep{clark13}, which is currently observed in the fast phase
of Case~A mass transfer. We speculate that the system may then emerge in a similar configuration to
\object{W36} or \object{W53a}, which are directly comparable in $I$-band and X-ray luminosity. The
lack of a robust period for \object{W10} makes its state less certain; most potential periodicities
are greater than ten days, but shorter periods are not yet ruled out. If it has $P<10$~days then it
must also be close to Roche-lobe overflow and circularisation. 

At longer periods ($P>10$~days), all OB stars in our sample have non-zero eccentricity.
\object{W44}/\object{L} is the only longer-period object with a circular orbit, and is clearly an
interacting system that may represent the next evolutionary stage of the luminous B2~Ia supergiants
\object{W2a}, \object{W23a}, and \object{W71}. We examine \object{W44}/\object{L} further in
Ritchie~et~al.~(in~prep.). 

Finally, Figure~\ref{fig:Iper} demonstrates the selection effects discussed previously.
\object{W1030} is the only system with $P>100$~days, reflecting our lack of sensitivity to
longer-period systems, but there is also a general lack of lower-luminosity systems (i.e. $I$-band
magnitudes fainter than $\sim$13, corresponding to objects earlier than $\sim$O9.5~II) with
$P>10$~days. The low $S$/$N$, broad lines, and low semi-amplitude of such systems make direct
confirmation of binarity impossible with our current data. Similarly, there is a lack of very
luminous, short period systems. Such objects have already undergone binary interaction, and are now
represented in the binary-rich Wolf-Rayet population or as disrupted, post-supernova systems: the
early-mid B hypergiants \object{W5}, \object{W1049}, and \object{W1069} may be examples of the
latter evolutionary endpoint.  

\subsection{Future Observations}
Finally, we turn to future observations of Wd1. The HR21 mode of FLAMES works well for targets of
spectral type O9.5~II and later, and further observations would support the discovery of additional
long period and high-eccentricity systems, and would allow refinement of the initial solutions
presented here. \object{W10}, \object{W24}, \object{W52}, and \object{WR~M} are of particular
importance in this respect, as they mark the entry and exit points of the short-period binary
evolutionary channel in Wd1, while the potential eclipsing system \object{W1048} may place
constraints on the masses of systems entering the luminous B~supergiant phase. 

HR21 mode is less effective for the lower-luminosity O9~III population, and future programmes should
also target the He~I\lam7065 absorption line covered by FLAMES HR16 and LR6 modes. Although HR16
mode proved unsuitable for detecting RV variability in the WN population, spare fibres were
allocated to a few bright supergiants, including \object{W10} and \object{W56a}, and demonstrated
that high quality observations of He~I\lam7065 can be obtained. However, no fainter targets were
observed and the performance of this mode at low $S$/$N$ cannot be assessed. LR6 covers H$\alpha$
and He~I\lam\lam6678,7065, and approximately half of our sample was observed in this mode.
He~I\lam7065 remains well defined down to O9~III in many cases, and appears better-defined than
Pa-11 in the faintest targets. Some confirmed binaries discussed in Section~\ref{sec:individual}
were observed twice in LR6 mode, and observations of \object{W1063} (O9~III) demonstrate that
He~I\lam7065 may be used to measure RVs in the faintest objects in our target list. Our only
observations of \object{W36} and \object{W1021} are in the $I$-band, but we note that \cite{kb12}
successfully measured RVs for these objects from He~I\lam7065. However, the line is broad and weak
in the majority of objects classified O9-9.5~III~bin? or O+O? in Paper~III, suggesting blending is
still present. Nevertheless, LR6 and HR16 modes appear a promising route forward for future
observations of Wd1, although their performance requires further assessment. 

\section{Conclusions}
\label{sec:conclusions}
In this paper, we have presented multi-epoch observations of more than 100 OB and Wolf-Rayet stars
in the massive Galactic cluster Westerlund~1. We summarise our findings as follows:
\begin{enumerate}
  \item We find OB binaries with spectral types that range from B2~Ia to O9~III, with the 173.9~day
  eccentric system \object{W1030} representing the first longer-period binary detected in the
  cluster. \object{W10} and \object{W53} are confirmed as SB2s, although both require follow-up
  observations to determine robust orbital solutions, while the simultaneous presence of He~I and
  C~III in the spectrum of \object{W23a} is inconsistent with an isolated star and implies a
  B2~Ia+O9~I binary. Several other objects display signs of composite spectra, most commonly in the
  anomalous strength of C~III~\lam8500. 
  \item \object{W232} appears unique in displaying two distinct and stable periodicities, with one
  exactly 11~times the other, indicating the presence of tidally-excited oscillations in a
  $9.98$~day eccentric system: it may be a massive `heartbeat' star \citep{fuller17}. This object is
  unusually well sampled, and other examples may be found in future observations, especially if
  contemporaneous RV and photometric data become available.  
  \item Twelve candidate OB binaries are identified. Five display $\Delta$RV$>30$\kms, but only
  limited observations are available, precluding further analysis: additional data will likely
  confirm these systems as binary. In addition, a number of other poorly-sampled objects are not
  currently identified as candidate binaries, but show secondary indicators of binarity or have mean
  RVs offset from the cluster mean, suggesting incomplete sampling of the RV curve. Additional
  observations of these objects may also confirm their (candidate) binary nature.   
  \item Pulsational instability appears to grow from O9~II to B2~Ia, and a lower RV cut might be
  appropriate for luminosity class II-III. We have not examined a varying RV cut in this paper,
  which would require a better understanding of pulsational instability in these objects, but note
  that adopting a $\Delta$RV$>20$\kms cut for our O9.5~II~-~O9~III targets would increase the number
  of candidate binaries by $\sim$50\%.  
  \item Many of the faintest targets in our survey display broad, flat-bottomed Paschen series line
  profiles, with Paper~III proposing that these systems are heavily blended double-lined systems. We
  find evidence to support this hypothesis, but in most cases cannot confirm the binary nature of
  these systems: only a few objects with this broad-lined morphology display RV changes large enough
  for conclusive classification as spectroscopic binaries.  
  \item A number of objects have been identified as strong binary candidates in previous X-ray
  \citep{clark08} or radio \citep{dougherty,fenech} observations, but could not be confirmed here.
  In some cases this may reflect insufficient observations (e.g. the strong X-ray source
  \object{W27}), but may also be a result of unfavourable inclination; in a large sample, we would
  expect some objects with very strong secondary indicators of binarity to be viewed at an angle
  that results in negligible RV variability. 
  \item Amongst the Wolf-Rayet population, we find an orbital solution for the 53.95~day WN9h:
  binary \object{W44}/\object{WR~L} and detect large RV changes in the WC9d star \object{WR~M},
  extending the list of Wolf-Rayet stars with massive, unseen companions in Wd1.
  \object{W44}/\object{WR~L} is the first longer-period Wolf-Rayet binary identified in the cluster,
  and likely represents a different evolutionary pathway to the $P<10$~day WR+O? systems
  \object{W72}/\object{WR~A}, \object{WR~B}, and \object{W239}/\object{WR~F} \citep{bonanos07,
  clark11}. 
  \item Almost all luminous binary systems with $P<10$~days appear to be interacting. The hard X-ray
  source \object{W24} (O9~Iab+O9~I?) has a candidate period of 6.595~days but shows no apparent sign
  of binary interaction; if confirmed, this system must be near the onset of Roche-lobe overflow,
  and may be a direct precursor to the X-ray luminous sgB[e] star \object{W9} \citep{clark13}. We do
  not find any very luminous binaries ($I<13$) with $P<5$~days.
  \item All OB binaries with $P>10$~days have non-zero eccentricity and appear to be pre-interaction
  systems: the B2~Ia+O9~I binary \object{W23a} may represent an evolutionary precursor to
  \object{W44}/\object{WR~L}. The apparent lack of X-ray luminous systems with longer periods is
  likely a selection effect: as discussed in Paper~II, X-ray detections peak at O9--9.5~I, but as
  orbital periods increase such systems become increasingly difficult to confirm as binary. 
  \item Although it is too early to attempt a robust determination of the OB binary fraction in Wd1,
  our results provide direct confirmation that the cluster is binary-rich, as suggested by earlier
  multiwavelength studies \citep{crowther06,clark08}. A preliminary estimate suggests a binary
  fraction in excess of $\sim40\%$, consistent with other massive clusters \citep{sana11}. However,
  our survey is optimised for short-period systems, and our understanding of the population of
  longer-period binaries is very limited: future observations of Wd1 may therefore significantly
  raise this estimate. 
\end{enumerate}

\begin{acknowledgements}
  The authors thank the referee for helpful comments and suggestions that have clarified this paper.
  Based on observations collected at the European Southern Observatory under programme IDs ESO
  081.D-0324, 383.D-0633, 087.D-0440, and 091.D-0179. This research is partially supported by the
  Spanish Government Ministerio de Ciencia, Innovaci\'on y Universidades under grant
  PGC2018-093741-B-C21 (MICIU/AEI/FEDER, UE). I.N. is also supported by the Generalitat Valenciana
  through grant PROMETEO/2019/041. F.N. acknowledges financial support through Spanish grant
  PID2019-105552RB-C41 (MINECO/MCIU/AEI/FEDER) and from the Spanish State Research Agency (AEI)
  through the Unidad de Excelencia "Mar\'ia de Maeztu"-Centro de Astrobiolog\'ia (CSIC-INTA) project
  No. MDM-2017-0737. 
\end{acknowledgements}

\clearpage
\onecolumn

\longtab{2}{
\label{tab:fullresults}
\begin{longtable}{llcccccc}
  \caption{Summary of Radial Velocity observations. Binaries listed in Table~\ref{tab:results} are shown in \textbf{bold}, and candidate binaries listed in Table~\ref{tab:candidates} are shown in \textit{italics}.}\\
ID   & Spectral Type & Observations & P(null) & Mean RV & RV Range & Mean Error & $\sigma_\textrm{det}$ \\
     &               &              &         & (\kms)    & (\kms)     & (\kms)       & \\
\hline
\\
\endfirsthead
\caption{continued.}\\
ID   & Spectral Type & Observations & P(null) & Mean RV & RV Range & Mean Error & $\sigma_\textrm{det}$ \\
&               &              &         & (\kms)    & (\kms)     & (\kms)       & \\
\hline
\\
\endhead
\hline
\endfoot
W1            & O9.5~Iab         & $6$  & $0.000$ & $-59.26$ & $11.17$  & $1.91$ & $4.15$\\
\textbf{W2a}  & B2~Ia            & $14$ & $0.000$ & $-46.06$ & $42.12$  & $0.46$ & $55.91$\\
\textbf{W6a}  & B0.5~Iab         & $7$  & $0.000$ & $-38.90$ & $40.21$  & $9.72$ & $4.03$\\
\textbf{W6b}  & O9.5~III         & $9$  & $0.000$ & $-57.17$ & $84.38$  & $2.58$ & $20.92$\\
W8a           & F8~Ia$^+$        & $5$  & $0.000$ & $-46.55$ & $2.71$   & $0.24$ & $7.98$\\
W8b           & B1.5~Ia          & $11$ & $0.000$ & $-47.85$ & $16.18$  & $0.35$ & $39.26$\\
\textbf{W10}  & B0.5~I + O9.5~II & $10$ & $0.000$ & $-58.16$ & $149.69$ & $1.24$ & $68.39$\\
W12a          & F1~Ia$^+$        & $5$  & $0.000$ & $-44.71$ & $9.82$   & $0.65$ & $10.77$\\
\textit{W15}  & O9~Ib            & $6$  & $0.000$ & $-44.81$ & $19.46$  & $1.98$ & $5.66$\\
W16a          & A5~Ia$^+$        & $3$  & $0.000$ & $-40.60$ & $11.85$  & $0.56$ & $15.37$\\
\textit{W17}  & O9~Iab           & $7$  & $0.000$ & $-36.87$ & $22.22$  & $1.52$ & $10.08$\\
W21           & B0.5~Ia          & $11$ & $0.000$ & $-46.58$ & $24.36$  & $1.11$ & $16.75$\\
\textbf{W23a} & B2~Ia + O9~I     & $11$ & $0.000$ & $-50.78$ & $18.94$  & $0.52$ & $25.91$\\
\textbf{W24}  & O9~Iab + O9~I?   & $12$ & $0.000$ & $-51.14$ & $26.60$  & $1.76$ & $15.05$\\
\textit{W27}  & O7--8~Ia$^+$     & $4$  & $0.000$ & $-60.56$ & $13.20$  & $1.03$ & $5.58$\\
\textbf{W30a} & O4--6~Ia$^+$     & $16$ & $0.000$ & $-41.27$ & $34.17$  & $4.27$ & $5.37$\\
W37           & O9~Ib            & $6$  & $0.000$ & $-46.02$ & $21.96$  & $1.77$ & $8.12$\\
W38           & O9~Iab           & $6$  & $0.000$ & $-44.19$ & $21.01$  & $1.48$ & $9.93$\\
W41           & O9~Iab           & $3$  & $0.000$ & $-49.50$ & $20.97$  & $1.10$ & $8.92$\\
\textbf{W43a} & B0~Ia            & $11$ & $0.000$ & $-65.24$ & $145.47$ & $0.93$ & $117.76$\\
\textit{W47}  & O9.5~Iab         & $3$  & $0.000$ & $-58.08$ & $19.97$  & $1.75$ & $5.73$\\
W49           & B0~Iab           & $6$  & $0.000$ & $-38.11$ & $13.84$  & $1.29$ & $7.11$\\
W50b          & O9~III           & $7$  & $0.004$ & $-36.10$ & $13.73$  & $3.12$ & $3.05$\\
\textbf{W52}  & B1.5~Ia          & $5$  & $0.000$ & $-43.68$ & $37.96$  & $0.71$ & $37.27$\\
W55           & B0~Ia            & $18$ & $0.000$ & $-47.85$ & $21.43$  & $1.27$ & $11.10$\\
W56a          & B1.5~Ia          & $6$  & $0.000$ & $-55.37$ & $19.70$  & $0.63$ & $29.82$\\
W56b          & O9.5~Ib          & $6$  & $0.001$ & $-44.53$ & $12.82$  & $1.76$ & $4.85$\\
W57a          & B4~Ia            & $7$  & $0.000$ & $-53.13$ & $12.06$  & $0.32$ & $23.13$\\
W60           & B0~Iab           & $11$ & $0.000$ & $-40.71$ & $17.25$  & $1.37$ & $9.39$\\
\textit{W65}  & O9~Ib            & $9$  & $0.000$ & $-46.74$ & $18.83$  & $2.30$ & $6.27$\\
W70           & B3~Ia            & $6$  & $0.000$ & $-42.86$ & $17.73$  & $0.38$ & $31.46$\\
\textbf{W71}  & B2.5~Ia          & $18$ & $0.000$ & $-44.89$ & $29.37$  & $0.50$ & $45.95$\\
W74           & O9.5~Iab         & $6$  & $0.000$ & $-45.90$ & $11.06$  & $1.24$ & $5.47$\\
W75$^a$       & M4~Ia            & $4$  & $0.238$ & $-50.02$ & $1.39$   & $0.51$ & -\\
W78           & B1~Ia            & $11$ & $0.000$ & $-45.99$ & $20.49$  & $0.61$ & $23.44$\\
W84           & O9.5~Ib          & $14$ & $0.000$ & $-49.40$ & $19.84$  & $2.09$ & $5.44$\\
W86           & O9.5~Ib          & $6$  & $0.010$ & $-50.94$ & $8.63$   & $1.85$ & $3.51$\\
W228b         & O9~Ib            & $6$  & $0.000$ & $-45.21$ & $22.24$  & $2.68$ & $6.46$\\
\textbf{W232} & B0~Iab           & $36$ & $0.000$ & $-48.95$ & $50.01$  & $1.76$ & $20.42$\\
W237$^a$      & M3~Ia            & $8$  & $0.000$ & $-47.86$ & $2.42$   & $0.21$ & -\\
W238          & B1~Iab           & $7$  & $0.000$ & $-43.46$ & $17.54$  & $0.84$ & $14.39$\\
W243$^b$      & A3~Ia$^+$        & $20$ & $0.000$ & $-34.08$ & $24.33$  & $0.23$ & $76.46$\\
W265          & F1--5~Ia$^+$     & $19$ & $0.000$ & $-49.24$ & $14.05$  & $0.35$ & $27.16$\\
W373          & B0~Iab           & $11$ & $0.000$ & $-47.29$ & $17.46$  & $0.97$ & $11.18$\\
\textit{W1001}& O + O?           & $6$  & $0.000$ & $-42.43$ & $25.05$  & $3.36$ & $4.60$\\
\textbf{W1002}& O9--9.5~II + O?  & $10$ & $0.000$ & $-58.12$ & $66.00$  & $5.57$ & $8.43$\\
W1003         & O9--9.5~bin?     & $6$  & $0.000$ & $-45.63$ & $20.95$  & $3.48$ & $4.09$\\
W1005         & B0~Iab           & $15$ & $0.000$ & $-34.61$ & $22.40$  & $1.44$ & $10.15$\\
\textit{W1006}& O9--9.5~III~bin? & $8$  & $0.000$ & $-54.32$ & $27.82$  & $3.02$ & $6.35$\\
W1007         & O9--9.5~III      & $6$  & $0.024$ & $-46.13$ & $7.08$   & $1.86$ & $2.45$\\
W1008         & O9.5~II          & $6$  & $0.333$ & $-47.05$ & $7.28$   & $2.77$ & $2.05$\\
W1009         & B0~Ib            & $6$  & $0.000$ & $-40.80$ & $6.02$   & $1.92$ & $2.53$\\
W1010         & O + O?           & $6$  & $0.069$ & $-53.27$ & $36.65$  & $7.73$ & $2.66$\\
W1011         & O + O?           & $7$  & $0.005$ & $-48.11$ & $22.19$  & $4.18$ & $3.78$\\
W1012         & O9--9.5~III~bin? & $6$  & $0.001$ & $-40.49$ & $25.00$  & $5.18$ & $4.13$\\
\textbf{W1013}& O + O?           & $6$  & $0.000$ & $-57.40$ & $98.82$  & $7.32$ & $6.86$\\
W1014         & O9--9.5~III~bin? & $6$  & $0.002$ & $-41.18$ & $19.52$  & $4.60$ & $3.37$\\
W1015         & O9~III           & $6$  & $0.006$ & $-40.19$ & $14.73$  & $2.77$ & $3.72$\\
W1016         & O9--9.5~III~bin? & $6$  & $0.012$ & $-44.21$ & $14.64$  & $3.89$ & $3.04$\\
\textit{W1017}& O9--9.5~III~bin? & $9$  & $0.000$ & $-53.76$ & $25.26$  & $4.00$ & $4.08$\\
\textit{W1018}& O9.5~Iab         & $4$  & $0.000$ & $-46.86$ & $33.16$  & $3.99$ & $5.88$\\
W1019         & O9--9.5~III~bin? & $7$  & $0.002$ & $-47.45$ & $26.30$  & $4.36$ & $4.18$\\
W1020         & O9--9.5 + O?     & $6$  & $0.003$ & $-44.49$ & $14.68$  & $2.82$ & $3.75$\\
\textbf{W1022}& O9.5~II          & $10$ & $0.000$ & $-34.87$ & $122.69$ & $2.83$ & $33.19$ \\
W1023         & O9~III           & $7$  & $0.059$ & $-49.94$ & $11.47$  & $2.42$ & $2.84$\\
W1024         & O9.5--B0~Ib      & $6$  & $0.001$ & $-43.47$ & $9.41$   & $1.70$ & $4.15$\\
W1025         & O + O?           & $6$  & $0.186$ & $-41.22$ & $16.15$  & $5.90$ & $1.52$\\ 
\textit{W1026}& O9--9.5~III      & $9$  & $0.000$ & $-37.85$ & $39.76$  & $4.26$ & $5.25$\\
\textbf{W1027}& O9.5~Iab         & $4$  & $0.000$ & $-40.97$ & $50.22$  & $2.97$ & $13.16$\\
\textit{W1028}& O9--9.5~III~bin? & $9$  & $0.000$ & $-48.97$ & $27.67$  & $4.40$ & $4.76$\\
\textit{W1029}& O9--9.5~III~bin? & $6$  & $0.000$ & $-43.27$ & $25.67$  & $2.82$ & $6.81$\\
\textbf{W1030}& O9.5~Ib          & $22$ & $0.000$ & $-46.77$ & $34.81$  & $1.72$ & $13.02$\\
\textit{W1031}& O9~III           & $7$  & $0.000$ & $-39.70$ & $20.05$  & $2.58$ & $5.13$\\
W1032         & O9--9.5~III~bin? & $7$  & $0.001$ & $-37.88$ & $28.60$  & $4.80$ & $4.34$\\
\textit{W1033}& O9--9.5~I--III   & $6$  & $0.161$ & $-43.70$ & $7.92$   & $2.29$ & $2.42$\\
W1034         & O9.5~Iab         & $1$  & -       & $-49.06$ & -        & $2.16$ & -\\
W1035         & O9--9.5~III~bin? & $11$ & $0.000$ & $-39.57$ & $24.56$  & $4.12$ & $3.90$\\
W1036         & O9.5~Ib          & $13$ & $0.266$ & $-30.89$ & $9.92$   & $2.50$ & $2.85$\\
W1037         & O9.5~II          & $6$  & $0.295$ & $-39.44$ & $8.78$   & $3.46$ & $1.91$\\
W1038         & O9~III           & $6$  & $0.000$ & $-40.47$ & $13.74$  & $2.57$ & $4.24$\\
W1039         & B1~Ia            & $6$  & $0.000$ & $-45.52$ & $10.77$  & $0.70$ & $11.93$\\
\textit{W1040}& O9.5~Iab         & $6$  & $0.000$ & $-49.25$ & $13.72$  & $1.55$ & $5.82$\\
\textit{W1041}& O9.5~Iab bin?    & $5$  & $0.285$ & $-46.80$ & $19.79$  & $6.90$ & $2.03$\\
\textit{W1042}& O9.5~II          & $8$  & $0.000$ & $-36.55$ & $47.96$  & $2.80$ & $12.55$\\
\textit{W1043}& O9.5~II--III     & $6$  & $0.000$ & $-49.67$ & $31.40$  & $3.24$ & $7.86$\\
W1044         & O9.5~III?        & $7$  & $0.151$ & $-38.47$ & $12.14$  & $4.48$ & $2.11$\\
W1045         & O9.5~II          & $6$  & $0.008$ & $-40.27$ & $11.29$  & $2.71$ & $3.05$\\
\textit{W1046}& O + O?           & $6$  & $0.057$ & $-44.01$ & $12.30$  & $3.03$ & $2.74$\\
\textit{W1047}& O9.5~II          & $10$ & $0.000$ & $-46.65$ & $35.19$  & $1.84$ & $12.76$\\
\textbf{W1048}& O9.5~Ib          & $9$  & $0.000$ & $-82.16$ & $151.24$ & $2.23$ & $43.23$\\
W1049         & B1--2~Ia$^+$     & $12$ & $0.000$ & $-59.67$ & $22.43$  & $0.69$ & $24.18$\\
\textbf{W1050}& O9.5~II          & $11$ & $0.000$ & $-56.36$ & $54.26$  & $2.65$ & $14.81$\\
W1051         & O9~III           & $6$  & $0.003$ & $-39.32$ & $8.64$   & $1.77$ & $3.67$\\
W1052         & O9~III           & $3$  & $0.741$ & $-41.00$ & $5.58$   & $5.06$ & $0.71$\\
W1053         & B0~Ib            & $3$  & $0.000$ & $-46.99$ & $10.15$  & $0.97$ & $7.06$\\
W1054         & O9--9.5~bin?     & $6$  & $0.004$ & $-45.95$ & $11.10$  & $2.40$ & $3.61$\\ 
\textit{W1055}& B0~Ib            & $6$  & $0.000$ & $-35.66$ & $15.70$  & $1.40$ & $8.30$\\
\textbf{W1056}& O9.5~II          & $10$ & $0.000$ & $-44.65$ & $51.06$  & $2.27$ & $16.86$\\
W1057         & O9.5~Ib          & $9$  & $0.001$ & $-41.43$ & $19.76$  & $2.86$ & $5.18$\\
W1058         & O9~III           & $6$  & $0.000$ & $-45.75$ & $15.82$  & $2.57$ & $4.30$\\
W1059         & O9~III?          & $6$  & $0.445$ & $-39.62$ & $9.99$   & $4.23$ & $1.80$\\
\textbf{W1060}& O9.5~II          & $10$ & $0.000$ & $-54.63$ & $56.92$  & $3.02$ & $11.40$\\
\textit{W1061}& O9--9.5~III~bin? & $10$ & $0.000$ & $-37.67$ & $26.24$  & $2.96$ & $6.77$\\
W1062         & O + O?           & $6$  & $0.019$ & $-45.99$ & $14.70$  & $4.10$ & $2.48$\\
\textbf{W1063}& O9~III           & $10$ & $0.000$ & $-51.01$ & $76.77$  & $2.70$ & $18.68$\\
W1064         & O9.5~Iab         & $3$  & $0.005$ & $-51.20$ & $26.74$  & $5.79$ & $3.27$\\
\textbf{W1065}& B0~Ib            & $27$ & $0.000$ & $-35.53$ & $79.47$  & $1.02$ & $44.60$\\
\textit{W1066}& O9~III           & $6$  & $0.000$ & $-37.20$ & $25.22$  & $2.49$ & $7.09$\\ 
\textbf{W1067}& B0~Iab           & $17$ & $0.000$ & $-47.75$ & $29.02$  & $0.89$ & $25.53$\\
W1068         & B0~Ib            & $6$  & $0.000$ & $-30.83$ & $22.17$  & $2.20$ & $8.91$\\
W1069         & B5~Ia$^+$        & $12$ & $0.000$ & $-43.23$ & $13.67$  & $0.88$ & $10.39$\\
\hline\\
\textbf{W13}$^c$        & WNVL~+~OB & $11$ & $0.000$ & $-48.20$ & $292.0$  & $9.45$ & $32.80$\\
\textbf{WR~C}$^d$       & WC9d      & $4$  & $0.084$ & $-67.10$ & $7.79$   & $3.23$ & $1.39$\\
WR~E (W241)$^d$         & WC9       & $4$  & $0.000$ & $-78.74$ & $17.66$  & $2.16$ & $7.73$\\
\textbf{WR~F (W239)}$^d$& WC9d      & $11$ & $0.000$ & $-60.50$ & $70.40$  & $3.51$ & $14.22$\\
\textbf{WR~L (W44)}$^e$ & WN9h:~+~O?& $6$  & $0.000$ & $-63.73$ & $150.65$ & $2.82$ & $62.16$\\
\textbf{WR~M (W66)}$^d$ & WC9d~+~O? & $4$  & $0.000$ & $-63.41$ & $102.71$ & $2.58$ & $28.18$\\
WR~S (W5)               & WN10--11h & $6$  & $0.000$ & $-97.65$ & $7.82$   & $0.69$ & $7.50$\\
\textbf{WR~T}$^d$       & WC9d      & $3$  & $0.000$ & $-57.28$ & $22.33$  & $2.18$ & $6.12$\\
\\
\end{longtable}
{Notes:\\
$^{a}$RVs measured from Ca~II\lamlam8542,8662. Epoch-to-Epoch variability $<1$\kms.\\  
$^{b}$RVs measured from Ca~II\lamlam8912,8927 (see \citealt{ritchie09b}).\\
$^c$Values given for the more massive absorption-line components \citep{ritchie10}.\\
$^d$RVs measured from the C~III~\lam8664 emission line (see \citealt{clark11}).\\
$^e$RVs measured from He~I~\lam8845. Values presented for 2011 and 2013 FLAMES observations only. 
}}
\clearpage
\twocolumn

\end{document}